\documentclass{aa}  
\usepackage{graphicx}       
\usepackage{natbib}
\usepackage{longtable}
\usepackage{supertabular}

\newcommand{\Ha} {H$\alpha$}
\newcommand{\Hb} {H$\beta$}
\newcommand{\HII}{\textsc{H\,ii\ }}

\newcommand{\Mo} {$M_{\odot}$}

\newcommand{\NIR} {\emph{NIR}}

\newcommand{\UV}{\emph{UV}}

\newcommand{\STARBURST}{{\sc Starburst~99}}
\newcommand{\PEGASE}{{\sc Pegase.2}}
\newcommand{\CLOUDY}{{\sc Cloudy}}
\newcommand{\MAPPINGS}{{\sc Mappings}}

\newcommand{\tableline}{\hline}

\newcommand{\Te}{$T_{\rm e}$}

\newcommand{\abox}{12+log(O/H)}

\newcommand{\nodata}{...}
\begin{document}
   \title{Massive star formation in Wolf-Rayet galaxies\thanks{Based on observations made with NOT (Nordic Optical Telescope), INT (Isaac Newton 
Telescope) and WHT (William Herschel Telescope) operated on the island of La Palma jointly by Denmark, Finland, Iceland, Norway and Sweden (NOT) or 
the Isaac Newton Group (INT, WHT) in the Spanish Observatorio del Roque de Los Muchachos of the Instituto de Astrof\'\i sica de Canarias. 
Based on observations made at the Centro Astron\'omico Hispano Alem\'an (CAHA) at Calar Alto, 
operated by the Max-Planck Institut f\"ur Astronomie and the Instituto de Astrof\'{\i}sica de Andaluc\'{\i}a (CSIC).
}}

  \subtitle{IV b. Using empirical calibrations to compute the oxygen abundance}

   \author{\'Angel R. L\'opez-S\'anchez
          \inst{1,2}
		  \and
		  C\'esar Esteban\inst{2,3}
          }

   \offprints{\'Angel R. L\'opez-S\'anchez, \email{Angel.Lopez-Sanchez@csiro.au}}

\institute{CSIRO Astronomy \& Space Science / Australia Telescope National Facility, PO\,BOX\,76, Epping, NSW\,1710, Australia \and Instituto de 
Astrof{\'\i}sica de Canarias, C/ V\'{\i}a L\'actea S/N, E-38200, La Laguna, Tenerife, Spain \and Departamento de Astrof\'{\i}sica de la Universidad 
de La Laguna, E-38071, La Laguna, Tenerife, Spain}


   \date{Submitted to astro-ph on 29 April 2010}

 
  \abstract
   {We have performed a comprehensive multiwavelength analysis of a sample of 20 starburst galaxies that show  a substantial
population of very young massive stars, most of them classified as Wolf-Rayet (WR) galaxies. }
{We have analysed optical/\NIR\ colours, 
physical and chemical properties of the ionized gas, stellar, gas and dust content, star-formation rate and interaction degree 
(among many other galaxy properties) of our galaxy sample using multi-wavelength data. We compile 41 independent
star-forming regions  --with oxygen abundances between \mbox{\abox= 7.58} and 8.75--, of which 31 have a direct estimate 
of the electron temperature of the ionized gas. }
{This paper, only submitted to astro-ph, compiles the most common empirical
calibrations to the oxygen abundance, and
presents the comparison between the chemical abundances derived in these galaxies using the direct method
with those obtained through empirical calibrations, as it is published in L\'opez-S\'anchez \& Esteban (2010b).}
  {We find that
  (i) the Pilyugin method 
  (Pilyugin 2001a,b; Pilyugin \& Thuan 2005), 
which considers the $R_{23}$ and the $P$ parameters, 
is the best suited empirical calibration for these star-forming galaxies, (ii) the relations
between the oxygen abundance and the $N_2$ or the $O_3N_2$ parameters 
provided by Pettini \& Pagel (2004) give acceptable results for objects with \abox$>$8.0, and (iii) the results provided by empirical calibrations 
based on photoionization models 
(McGaugh, 1991; Kewley \& Dopita, 2002; Kobulnicky \& Kewley, 2004) 
are systematically 0.2 -- 0.3 dex higher than the values derived from the direct method. 
These differences are of the same order that the abundance discrepancy found between 
recombination and collisionally excited lines. This may suggest the existence of temperature 
fluctuations in the ionized gas, as exists in Galactic and other extragalactic \HII regions. }
{All these results are included in the paper \emph{Massive Star Formation in Wolf-Rayet galaxies IV. Colours, chemical-composition analysis and metallicity-luminosity relations}, L\'opez-S\'anchez \& Esteban (2010b), A\&A, in press (Sect.~4.4 and Appendix~A). Please, if this information is used, reference that paper and NOT this document, which have been only submitted to astro-ph to emphasize these results.}
 

\titlerunning{Massive star formation in Wolf-Rayet galaxies IVb: Empirical calibrations }

\authorrunning{L\'opez-S\'anchez \& Esteban}

   \keywords{galaxies: starburst --- galaxies: dwarf --- galaxies: abundances --- stars: Wolf-Rayet}
   \maketitle
%

\section{Introduction}

The knowledge of the chemical composition of galaxies, in particular of dwarf galaxies, is vital for understanding their evolution, star formation 
history, stellar nucleosynthesis, the importance of gas inflow and outflow, and the enrichment of the intergalactic medium. Indeed, metallicity is a 
key ingredient for modelling galaxy properties, because it determines \UV, optical and \NIR\ \mbox{colours} at a given age (i.e., Leitherer et al. 
1999), nucleosynthetic yields (e.g., Woosley \& Weaver 1995), the dust-to-gas ratio (e.g., Hirashita et al 2001), the shape of the interstellar 
extinction curve (e.g., Piovan et al. 2006), or even the properties of the Wolf-Rayet stars \citep{Crowther07}.

The most robust method to derive the metallicity in star-forming and starburst galaxies is via the estimate of metal abundances and abundance 
ratios, in particular through the determination of the gas-phase oxygen abundance and the nitrogen-to-oxygen ratio. The relationships between current 
metallicity and other galaxy parameters, such as colours, luminosity, neutral gas content, star-formation rate, extinction or total mass, constrain
galaxy-evolution models and give clues about the current stage of a galaxy. 
For example, is still debated whether massive star formation results in the instantaneous enrichment of the interstellar medium of a dwarf galaxy, or if 
the bulk of the newly synthesized heavy elements must cool before becoming part of the interstellar medium (ISM) that eventually will form the next 
generation of stars. Accurate oxygen abundance measurements of several \HII regions within a dwarf galaxy will increase the understanding of its 
chemical enrichment and mixing of enriched material.

\onecolumn

Furthermore, today it is the metallicity (which reflects the gas reprocessed by stars and any exchange of gas between the galaxy and its environment) and not  
the stellar mass (which reflects the amount of gas locked up into stars) of a galaxy the main problem to get a proper metallicity-luminosity relation, 
so that different methods involving 
direct estimates of the oxygen abundance, empirical calibrations using bright emission-line ratios or theoretical methods based on photoionization 
models yield very different values (i.e., Yin et al. 2007; Kewley \& Elisson, 2008).

Hence precise photometric and spectroscopic data, including a detailed analysis of each particular galaxy that allows conclusions about its nature, are 
crucial to address these issues. We performed such a detailed photometric and spectroscopic study in a sample of strong star-forming galaxies, 
many of them previously classified as dwarf galaxies. The majority of these objects are  
Wolf-Rayet (WR) galaxies,
a very inhomogeneous class of star-forming objects which share at least
an ongoing or recent star formation event that has produced stars sufficiently massive
to evolve into the WR stage \citep{SCP99}. 
The main aim of our study of the formation of 
massive stars in starburst galaxies and the role that the interactions with or between dwarf galaxies and/or low surface brightness objects have in 
its triggering mechanism. In Paper~I \citep{LSE08} we described the motivation of this work, compiled the list of the 20 analysed 
WR galaxies (Table~1 of Paper~I), the majority of them showing several sub-regions or objects within or surrounding them, and presented the results 
of the optical/\NIR\ broad-band and \Ha\ photometry. In Paper~II \citep{LSE09} we presented the results of the analysis of the intermediate 
resolution long-slit spectroscopy of 16 WR galaxies of our sample -- the results for the other four galaxies were published separately, see \citet{LSER04a,LSER04b,LSEGR06,LSEGRPR07}. 
In many cases, two or more slit positions were used 
to analyse the most interesting zones, knots or morphological structures belonging to each galaxy or even surrounding objects. 
Paper~III \citep{LSE10a} presented the analysis of the O and WR stellar populations within these galaxies. 
Paper~IV \citep{LSE10b} globally compile and analyse the optical/\NIR\ photometric data and study the physical
and chemical properties of the ionized gas within our galaxy sample. The results shown in this paper haven been already published in Paper~IV.
The final paper of the series (Paper~V; L\'opez-S\'anchez 2010) compiles the properties derived with data from other wavelengths (UV, FIR, radio, and X-ray) 
and complete a global analysis of all available multiwavelength data of our WR galaxy sample. We have produced the most comprehensive data 
set of these galaxies so far, involving multiwavelength results and analysed according to the same procedures.

\section{Empirical calibrations of the oxygen abundance\label{empiricalcalibrations}}

When the spectrum of an extragalactic \HII region does not show the [\ion{O}{iii}] $\lambda$4363 emission line or other auroral lines that can be 
used to derive \Te, the so-called \emph{empirical calibrations} are applied to get a rough estimation of its metallicity.
Empirical calibrations are inspired partly by photo-ionization models and partly by observational trends of line strengths with galactocentric 
distance in gas-rich spirals, which are believed to be due to a radial abundance gradient with abundances decreasing outwards. In extragalactic 
objects, the usefulness of the empirical methods goes beyond the derivation of abundance gradients in spirals \citep{P04}, as these methods find 
application in chemical abundance studies of a variety of objects, including low-surface brightness galaxies \citep{deNaray04} and star-forming 
galaxies at intermediate and high redshift, where the advent of 8--10 m class telescopes has made it possible to extend observations (e.g., Teplitz et 
al. 2000, Pettini et al. 2001; Kobulnicky et al 2003; Lilly, Carollo \& Stockton 2003; Steidel et al. 2004; Kobulnicky \& Kewley 2004; Erb et al. 
2006).

\begin{figure}[t!]
\centering
\includegraphics[angle=270,width=0.7\linewidth]{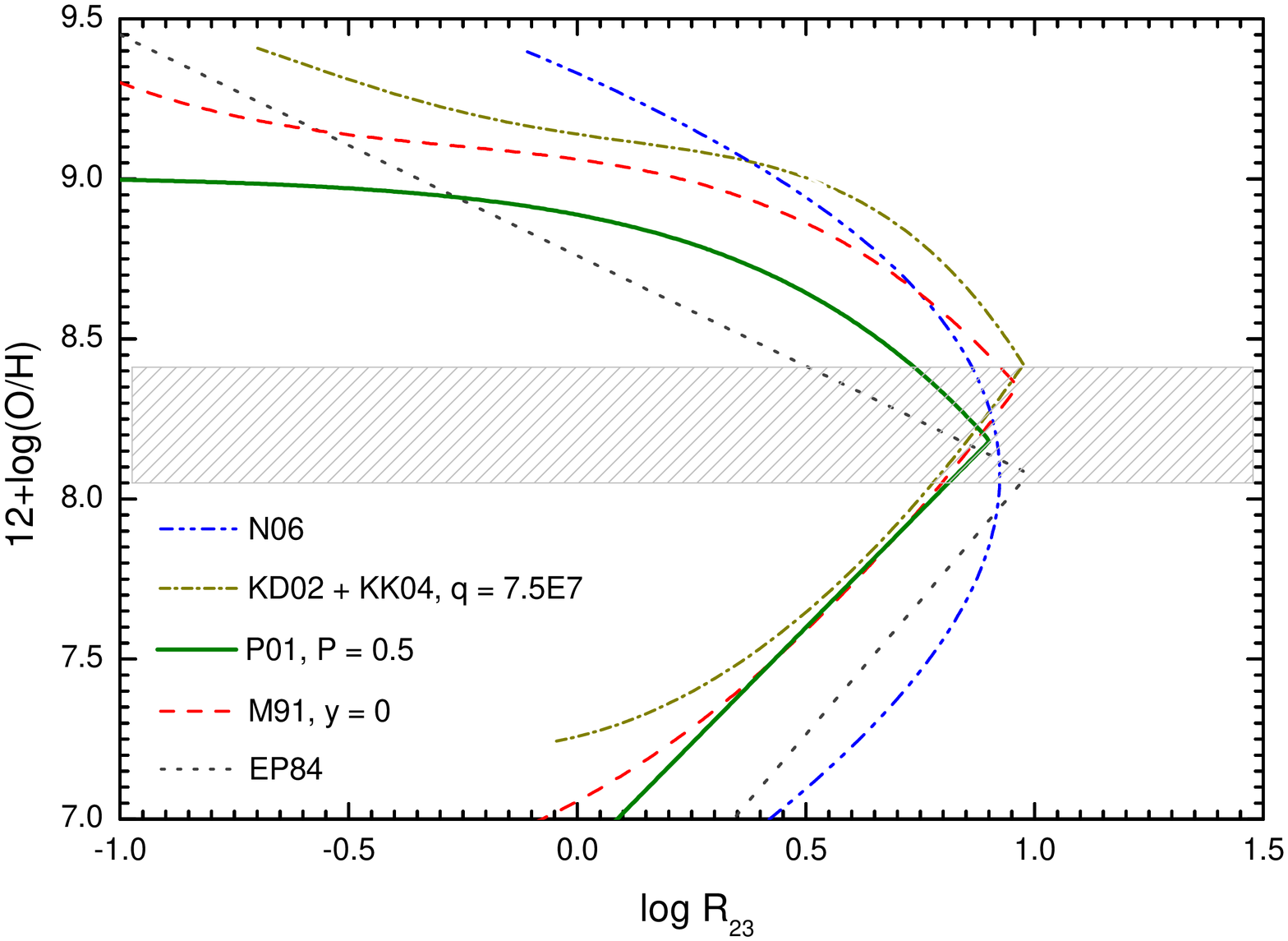}   
\caption{\footnotesize{Empirical calibrations of oxygen abundance using the $R_{23}$ parameter. Note that they are bi-valuated. The dashed zone 
indicates the region with higher uncertainties in O/H. The empirical calibrations plotted in the figure are: EP94: Edmund \& Pagel (1984); M91: 
McGaugh (1991) using $y$=0 ($R_2=R_3$); 
P01: Pilyugin (2001) using $P=0.5$ ($R_2=R_3$); (KD02+KK04): Kewley \& Dopita (2002) using the formulation of Kobulnicky \& Kewley (2004) assuming 
$q=7.5\times 10^7$ cm s$^{-1}$; N06: Nagao et al. (2006) using their cubic fit to $R_{23}$.}}
\label{cempiricas}
\end{figure}

As the brightest metallic lines observed in spectra of \HII regions are those involving oxygen, this element has been extensively used to get a 
suitable empirical calibration. Oxygen abundance is important as one of the fundamental characteristics of a galaxy: its radial 
distribution is combined with radial distributions of gas and star surface mass densities to constrain models of chemical evolution. Parameters 
defined in empirical calibrations evolving bright oxygen lines are
\begin{eqnarray}
R_3=  \frac{I([{\rm O\, III}]) \lambda 4959+I([{\rm O\, III}]) \lambda 5007}{\rm H\beta}, \\
R_2=  \frac{I([{\rm O\, II}]) \lambda 3727}{\rm H\beta}, \\
R_{23} =  R_3 + R_2, \\
P =  \frac{R_3}{R_{23}}, \\
y =  \log \frac{R_3}{R_2} = \log \frac{1}{P^{-1}-1}.
\end{eqnarray}   
\citet{Jensen76} presented the first exploration in this method considering the $R_3$ index, which considers the [\ion{O}{iii}] 
$\lambda\lambda$4959,5007 emission lines. However, were \citet{Pag79} who introduced the most widely used abundance indicator, the $R_{23}$ index, 
which also included the bright [\ion{O}{ii}] $\lambda$3727 emission line. Since then, many studies have been performed to refine the 
calibration of $R_{23}$ \citep{EP84,MRS85,DE86,Torres-Peimbert89,McGaugh91,ZKH94,P00,P01a,P01b,KD02,KK04,PT05,Nagao06}.
The most successful are the calibrations of \citet{McGaugh91} and \citet{KD02}, which are based on photoionization models, and the empirical 
relations provided by \citet{P01a,P01b} and \citet{PT05}. Both kinds of calibrations improve the accuracy by making use of the 
[\ion{O}{iii}]/[\ion{O}{ii}] ratio as ionization parameter, which accounts for the large scatter found in the $R_{23}$ versus oxygen abundance 
calibration, which is larger than observational errors \citep{KKP99}. Figure~\ref{cempiricas} shows the main empirical calibrations that use the 
$R_{23}$ parameter.

The main problem associated with the use of $R_{23}$ parameter is that it is {\bf bivaluated}, i.e., a single value of $R_{23}$ can be caused by two 
very different oxygen abundances. The reason of this behaviour is that the intensity of oxygen lines \emph{does not indefinitely increase} with 
metallicity. Thus, there are two \emph{branches} for each empirical calibration (see Fig.~\ref{cempiricas}): the \emph{low-metallicity} regime, 
with 12+log(O/H)$\leq$8.1, and the \emph{high-metallicity} regime, with 12+log(O/H)$\geq$8.4. That means that a very large fraction of the 
star-forming regions lie in the ill-defined turning zone around 12+log(O/H)$\simeq$8.20, where regions with the same $R_{23}$ value have oxygen 
abundances that differ by almost an order of magnitude. Hence, additional information, such as the [\ion{N}{ii}]/\Ha\ or the 
[\ion{O}{ii}]/[\ion{O}{iii}] ratios, is needed to break the degeneracy between the high and low branches (i.e., Kewley \& Dopita, 2002). 
Besides, the $R_{23}$ method requires that spectrophotometric data are corrected by reddening, which effect is crucial because [\ion{O}{ii}] and 
[\ion{O}{iii}] lines have a considerably separation in wavelength.

Here we list all empirical calibrations that were considered in this work, compiling the equations needed to derive the oxygen abundance from 
bright emission line ratios following every method.

{\bf Edmund \& Pagel (1984)}: Although the $R_{23}$ parameter was firstly proposed by \citet{Pag79}, the first empirical calibration was given by 
\citet{EP84},
\begin{eqnarray}
\nonumber 12+\log ({\rm O/H})_{up}= 8.76 -0.69 \log R_{23}, \\
12+\log ({\rm O/H})_{low} = 6.43 +1.67 \log R_{23},
\end{eqnarray}
with the limit between the lower and the upper branches at \abox$\sim$8.0.

{\bf McCall, Rybski \& Shields (1985)} 
presented an empirical calibration for oxygen abundance using the $R_{23}$ parameter, only valid for 12+log(O/H)$>$8.15. However, they did not give an 
analytic formulae but only listed it numerically (see their Table~15). The four-order polynomical fit for their values gives the following relation:
\begin{eqnarray}
12+\log ({\rm O/H})_{up} = & 9.32546  -0.360465x +0.203494x^2 
                                      &       +0.278702x^3   -1.36351x^4,
\end{eqnarray}
with $x=\log R_{23}$.

{\bf Zaritzky, Kennicutt \& Huchra (1994)} provided a simple analytic relation between oxygen abundance and $R_{23}$: 
\begin{eqnarray}
12+\log ({\rm O/H})_{up} =  9.265-0.33x  -0.202x^2  -0.207x^3 -0.333x^4.
\end{eqnarray}
Their formula is an average of three previous calibrations: \citet{EP84}, McCall et al. (1985) and Dopita \& Evans (1986). Following the authors, 
this calibration is only suitable for 12+log(O/H)$>$8.20, but perhaps a more realistic lower limit is 8.35.

{\bf McGaugh (1991)} calibrated the relationship between the $R_{23}$ ratio and gas-phase oxygen abundance using \HII region models derived from the 
photoionization code \CLOUDY\ \citep{Ferland98}. McGaugh's models include the effects of dust and variations in ionization parameter, $y$. 
\citet*{KKP99} give analytical expressions for the \citet{McGaugh91} calibration based on fits to photoionization models; the middle point between both 
branches is  \abox$\sim$8.4:
\begin{eqnarray}
12+\log ({\rm O/H})_{up} =  & 
7.056  +0.767x +0.602x^2  -y (0.29+0.332x-0.331x^2), \\
\nonumber 
12+\log ({\rm O/H})_{low} = & 9.061-0.2x-0.237x^2-0.305x^3-0.0283x^4 \\
& -y(0.0047-0.0221x-0.102x^2-0.0817x^3-0.00717x^4).
\end{eqnarray}

{\bf Pilyugin (2000)} found that the previous calibrations using the $R_{23}$ para\-meter had a systematic error depending on the hardness of the 
ionizing radiation, suggesting that the excitation parameter, $P$, is a good indicator of it. In several papers, Pilyugin performed a detailed 
analysis of the observational data combined with photoionization models to obtain empirical calibrations for the oxygen abundance. \citet{P00} 
confirmed the idea of \citet{McGaugh91} that the strong lines of [\ion{O}{ii}] and [\ion{O}{iii}] contain the necessary information for the determination 
of accurate abundances in low-metallicity (and may be also in high-metallicity) \HII regions. He used new observational data to propose a linear fit 
involving only the $R_{23}$ parameter,
\begin{eqnarray}
12+\log ({\rm O/H})_{up} = 9.50 - 1.40 \log R_{23},\\
12+\log ({\rm O/H})_{low} = 6.53 + 1.40 \log R_{23},
\end{eqnarray}
assuming a limit of \abox$\sim$8.0 between the two branches. This calibration is close to that given by \citet{EP84}; it has the same slope, but 
\citet{P00} is shifted towards lower abundances by around 0.07 dex. However, this new relation is not sufficient to explain the wide spread of 
observational data. Thus, {\bf Pilyugin (2001a)} give the following, more real and complex, calibration involving also the excitation parameter $P$:
\begin{eqnarray}
12+\log ({\rm O/H})_{up} =\frac{R_{23}+54.2+59.45P+7.31P^2}{6.01+6.71P+0.371P^2+0.243R_{23}}.
\end{eqnarray}
This is the so-called \emph{P-method}, which can be used in moderately high-metallicity \ion{H}{ii} regions (\abox$\geq$8.3). Pilyugin used two-zone 
models of \HII regions and assumed the $T_e$(\ion{O}{ii}) -- $T_e$(\ion{O}{iii}) relation from \citet{G92}. 
For the low metallicity branch, {\bf Pilyugin (2001b)} found that
\begin{eqnarray}
12+\log ({\rm O/H})_{low} = 6.35 + 1.45 \log R_{23} -1.74 \log P.
\end{eqnarray}
Pilyugin estimates that the precision of oxygen abundance determination 
with this method is around 0.1 dex.

{\bf Pilyugin \& Thuan (2005)} revisited these calibrations including more spectroscopic measurements of \HII regions in spiral and irregular 
galaxies with a measured intensity of the [\ion{O}{iii}] $\lambda$4363 line and recalibrate the relation between the oxygen abundance and the 
$R_{23}$ and $P$ parameters, yielding to:
\begin{eqnarray}
12+\log ({\rm O/H})_{low} = \frac{R_{23}+106.4+106.8P-3.40P^2}{17.72+6.60P+6.95P^2-0.302R_{23}},\\
12+\log ({\rm O/H})_{up} = \frac{R_{23}+726.1+842.2P+337.5P^2}{85.96+82.76P+43.98P^2+1.793R_{23}}.
\end{eqnarray}

{\bf Kewley \& Dopita (2002)} used a combination of stellar population synthesis and photoionization models to develop a set of ionization parameters 
and abundance diagnostic based only on the strong optical emission lines. Their \emph{optimal} method uses ratios of [\ion{N}{ii}], [\ion{O}{ii}], 
[\ion{O}{iii}], [\ion{S}{ii}], [\ion{S}{iii}] and Balmer lines, which is the full complement of strong nebular lines accessible from the ground. They 
also recommend procedures for the derivation of abundances in cases where only a subset of these lines is available.  \citet{KD02} models start with 
the assumption that $R_{23}$, and many of the other emission-line abundance diagnostics, also depends on the {\bf ionization parameter} $q \equiv 
c\times U$, that has units of \mbox{cm~s$^{-1}$.} They used the ste\-llar population synthesis codes \STARBURST\ \citep{L99,VL05} and \PEGASE\ 
\citep{PEGASE97} to generate the ionizing radiation field, assuming burst models at zero age with a Salpeter IMF and lower and upper mass limits of 
0.1 and 120 \Mo, respectively, with metallicities between 0.05 and 3 times solar. The ionizing radiation fields were input into the photoionization 
and shock code, \MAPPINGS\ \citep{SDopita93}, which includes self-consistent treatment of nebular and dust physics. \citet{KD02} previously used these 
models to simulate the emission-line spectra of \HII regions and starburst galaxies \citep{Do00}, and are completely described in their study.

{\bf Kobulnicky \& Kewley (2004)} gave a parameterization of the \citet{KD02} $R_{23}$ method with a form similar to that given by \citet{McGaugh91} 
calibration. 
\citet{KK04} presented an iterative scheme to resolve for both the ionization parameter $q$ and the oxygen abundance using only [\ion{O}{iii}], 
[\ion{O}{ii}] and \Hb\ lines. The parameterization they give for $q$ is
\begin{eqnarray}
\log(q)= \frac{32.81 - 1.153y^2 + \big[ 12+\log({\rm O/H}) \big] \big[ -3.396-0.025y+0.1444y^2 \big]}{4.603-0.3119y-0.163y^2 + \big[ 12+\log({\rm 
O/H}) \big] \big[ -0.48+0.0271y+0.02037y^2 \big]},
\end{eqnarray}
where $y=\log$([\ion{O}{iii}]/[\ion{O}{ii}]). This equation is only valid for ionization parameters between 5$\times$10$^6$ and 1.5$\times$10$^8$ cm 
s$^{-1}$. The oxygen abundance is parameterized by
\begin{eqnarray}
\nonumber 12+\log ({\rm O/H})_{up} =  9.72-0.777x-0.951x^2-0.072x^3-0.811x^4 -\log(q) \\
   \times (0.0737-0.0713x-0.141x^2+0.0373x^3-0.058x^4),\\
12+\log ({\rm O/H})_{low} = 9.40+4.65x-3.17x^2-\log(q)(0.272+0.547x-0.513x^2), 
\end{eqnarray}
being $x=\log R_{23}$. The first equation is valid for 12+log(O/H)$\geq$8.4, while the second for 12+log(O/H)$<$8.4. Typically, between two and three 
iterations are required to reach convergence. Following the authors, this parameterization should be regarded as an improved, implementation-friendly 
approach to be preferred over the tabulated $R_{23}$ coefficients given by \citet{KD02}.  

{\bf Nagao, Maiolino \& Marcani (2006)} did not consider any ionization parameter. They merely used data of a large sample of galaxies from the \emph{Sloan Digital Sky Survey} (SDSS; York et al. 2000)
 to derive a cubic fit to the relation between $R_{23}$ and the oxygen abundance,
\begin{eqnarray}
\log R_{23} = 1.2299 -4.1926y+1.0246y^2-0.063169y^3,
\end{eqnarray}
with $y$=\abox.

Besides $R_{23}$, additional parameters have been used to derive metallicities in star-forming galaxies.  Without other emission lines, the 
{\bf $N_2$ parameter}, which is defined by
\begin{eqnarray}
N_2 \equiv \log \frac{I([{\rm N\, II}]) \lambda 6583}{\rm H\alpha},
\end{eqnarray}
can be used  as a crude estimator of metallicity. However, we note that the [\ion{N}{ii}]/\Ha\ ratio is particularly sensitive to shock excitation or 
a hard radiation field from an AGN. The $N_2$ parameter was firstly suggested by \citet{SBCK94}, who gave a tentative calibration of 
the oxygen abundance using this parameter. This calibration has been revisited by \citet{vZee98,D02,PP04} and \citet*{Nagao06}.
The {\bf Denicol\'o et al. (2002)} calibration is
\begin{eqnarray}
12+\log ({\rm O/H}) = 9.12 + 0.73 N_2,
\end{eqnarray}
which considerably improves the previous relations because of the inclusion of an extensible sample of nearby extragalactic \HII regions. The 
uncertainty of this method is $\sim$0.2 dex because $N_2$ is sensitive to ionization and O/N variations, so strictly speaking it should be used mainly as an 
indicator of galaxy-wide abundances. \citet*{D02} also compared their method with photoionization models, concluding that the observed $N_2$ is 
consistent with nitrogen being a combination of both primary and secondary origin. 

\citet{PP04} revisited the relation between the $N_2$ parameter and the oxygen abundance including new data for the high- and low-metallicity regimen. 
They only considered those extragalactic \HII regions where the oxygen values are determined either via the $T_e$ method or with detailed photoionization 
modelling. The linear fit to their data is
\begin{eqnarray}
12+\log ({\rm O/H}) = 8.90 + 0.57 N_2,
\end{eqnarray}
which has both a lower slope and zero-point that the fit given by \citet*{D02}. A somewhat better relation is provided by a third-order polynomical fit 
of the form
\begin{eqnarray}
12+\log ({\rm O/H}) = 9.37 + 2.032 N_2 + 1.26 (N_2)^2 + 0.32 (N_2)^3,
\end{eqnarray}
valid in the range $-2.5 < N_2 < -0.3$. \citet*{Nagao06} also provided a relation between $N_2$ and the oxygen abundance, their cubic fit to their 
SDSS data yields
\begin{eqnarray}
\log N_2 = 96.641 -39.941y + 5.2227y^2 -0.22040y^3,
\end{eqnarray}  
with $y$=\abox.

{\bf Pettini \& Pagel (2004)} revived the O$_3$N$_2$ parameter, previously introduced by \citet{Alloin79} and defined by
\begin{eqnarray}
O_3N_2 \equiv \log\frac{[{\rm O\, III}]\ \lambda 5007/{\rm H\beta}}{[{\rm N\, II}]\ \lambda 6583/{\rm H\alpha}}.
\end{eqnarray} 
\citet{PP04} derived the following least-square linear fit to their data:
\begin{eqnarray}
12+\log({\rm O/H}) = 8.73-0.32 O_3N_2.
\end{eqnarray}
\citet*{Nagao06} also revisited this calibration and derived a cubic fit between the O$_3$N$_2$ parameter and the oxygen abundance,
\begin{eqnarray}
\log O_3N_2 = -232.18 + 84.423y -9.9330y^2 +0.37941y^3,
\end{eqnarray}  
with $y$=\abox.

Other important empirical calibrations that were not used in this study involve the $S_{23}$ parameter, introduced by \citet{Vil96} and 
revisited by \citet{Dia00,OS00} and \citet{PerezMontero05}. In the last years, bright emission line ratios such as  [\ion{Ar}{iii}]/[\ion{O}{iii}] 
and [\ion{S}{iii}]/[\ion{O}{iii}] \citep{Sta06} or [\ion{Ne}{iii}]/[\ion{O}{iii}] and [\ion{O}{iii}]/[\ion{O}{ii}]  \citep*{Nagao06} have been 
explored as indicators of the oxygen abundance in \HII regions and starburst galaxies. \citet{Peimbert07} suggested to use the oxygen recombination 
lines to get a more precise estimation of the oxygen abundance. Nowadays, there is still a lot of observational and theoretical work to do involving 
empirical calibrations (see recent review by Kewley \& Ellison 2008), but these methods should be used only for objects whose \HII regions have the 
same structural properties as those of the calibrating samples \citep{Stasinska09}.

\begin{table}[t!]
\centering
  \caption{\footnotesize{List of the parameters used to compute the oxygen abundance in all regions with a direct estimation of \Te\ using empirical calibrations.}}
  
  \label{abempirica1}
  \tiny
  \begin{tabular}{l@{\hspace{5pt}} l@{\hspace{5pt}}   c@{\hspace{5pt}}c@{\hspace{5pt}}c@{\hspace{5pt}}       
c@{\hspace{5pt}}c@{\hspace{5pt}}c@{\hspace{5pt}} }
  \tableline
   \noalign{\smallskip}
Region & $R_{23}$ & $P =R_3/R_{23} $ & $y$ =log($R_3/R_2$) & $N_2$ & $O_3N_2$ & $q_{KD02o}$$^a$ \\ 
\tableline
\noalign{\smallskip}   
       HCG~31~AC  &  5.42  &  0.571  &  0.125  &  0.104  &   1.349  &  3.76E+07  \\  
       HCG~31~B	  &  7.93  &  0.408  & -0.162  &  0.101  &   1.381  &  4.91E+07  \\  
       HCG~31~E   &  7.12  &  0.511  &  0.020  &  0.090  &   1.486  &  7.40E+07  \\  
      HCG~31~F1   &  8.91  &  0.819  &  0.656  &  0.034  &   2.201  &  5.78E+07  \\  
      HCG~31~F2   &  7.60  &  0.724  &  0.418  &  0.036  &   2.064  &  6.28E+07  \\  
       HCG~31~G   &  8.20  &  0.499  & -0.002  &  0.106  &   1.462  &  6.96E+07  \\  
	   Mkn~1199~C &  7.69  &  0.809  &  0.627  &  0.131  &   1.555  &  1.55E+08 \\
    Haro~15~A     &  9.73  &  0.884  &  0.881  &  0.027  &   2.378  &  8.55E+07  \\   
       Mkn~5~A1   &  7.58  &  0.748  &  0.473  &  0.051  &   1.915  &  6.96E+07  \\   
       Mkn~5~A2   &  8.19  &  0.702  &  0.372  &  0.049  &   1.944  &  1.72E+08  \\             
  IRAS~08208+2816~C&  7.77  &  0.793  &  0.583  &  0.129  &   1.558  &  8.55E+07  \\  
  POX~4   & 10.68  &  0.906  &  0.986  &  0.015  &   2.697  &  1.05E+08  \\   
        UM~420   &  6.45  &  0.649  &  0.268  &  0.099  &   1.497  &  4.81E+07  \\ 
 SBS~0926+606A   &  7.40  &  0.811  &  0.632  &  0.026  &   2.227  &  6.68E+07  \\   
SBS~0948+532   &  8.85  &  0.874  &  0.843  &  0.022  &   2.430  &  2.54E+08  \\   
SBS~1054+365   &  9.33  &  0.893  &  0.920  &  0.020  &   2.503  &  9.10E+07  \\  
SBS~1211+540   &  7.22  &  0.892  &  0.918  &  0.008  &   2.788  &  1.16E+08  \\  
SBS~1319+579A   &  9.92  &  0.908  &  0.996  &  0.014  &   2.671  &  1.05E+08  \\  
SBS~1319+579B   &  7.13  &  0.722  &  0.415  &  0.046  &   1.922  &  6.15E+07  \\  
SBS~1319+579C   &  7.11  &  0.710  &  0.389  &  0.052  &   1.860  &  5.91E+07  \\  
SBS~1415+437C   &  5.22  &  0.783  &  0.558  &  0.015  &   2.301  &  5.91E+07  \\  
SBS~1415+437A   &  4.86  &  0.810  &  0.629  &  0.012  &   2.370  &  5.44E+07  \\ 
   III~Zw~107~A    &  7.13  &  0.701  &  0.369  &  0.100  &   1.573  &  5.78E+07  \\ 
Tol~9~INT  &  4.58  &  0.689  &  0.345  &  0.252  &   0.973  &  4.16E+07  \\ 
      Tol~9~NOT   &  4.78  &  0.629  &  0.230  &  0.287  &   0.894  &  3.39E+07  \\ 
Tol~1457-262A   &  9.89  &  0.773  &  0.532  &  0.033  &   2.236  &  9.91E+07  \\ 
Tol~1457-262B   &  9.00  &  0.792  &  0.582  &  0.020  &   2.417  &  1.41E+08  \\ 
Tol~1457-262C   &  8.88  &  0.669  &  0.359  &  0.036  &   2.099  &  7.16E+07  \\ 
      ESO~566-8   &  5.17  &  0.505  &  0.008  &  0.414  &   0.693  &  3.19E+07  \\
     NGC~5253~A   &  9.20  &  0.851  &  0.756  &  0.102  &   1.754  &  6.82E+07  \\ 
     NGC~5253~B   &  9.38  &  0.856  &  0.775  &  0.086  &   1.841  &  7.11E+07  \\ 
     NGC~5253~C   &  8.03  &  0.773  &  0.532  &  0.041  &   2.046  &  2.60E+08  \\ 
     NGC~5253~D   &  7.67  &  0.527  &  0.048  &  0.079  &   1.582  &  7.72E+07  \\
\tableline
  \end{tabular}
      \begin{flushleft}
$^a$ Value derived for the $q$ parameter (in units of cm s$^{-1}$) obtained using the optimal calibration given by \citet{KD02}.
    \end{flushleft}

  \end{table}

\begin{table*}[t]
  \caption{\footnotesize{Results of the oxygen abundance, in the form \abox, for objects with a direct estimation of the metallicity, considering 
several empirical calibrations.}}
  \label{abempirica2}
    \tiny
  \begin{tabular}{l@{\hspace{2pt}} c@{\hspace{2pt}} c@{\hspace{2pt}} c@{\hspace{2pt}} c@{\hspace{2pt}} c@{\hspace{2pt}}  c@{\hspace{2pt}} 
c@{\hspace{2pt}} c@{\hspace{2pt}}  c@{\hspace{2pt}} c@{\hspace{2pt}} c@{\hspace{2pt}} c@{\hspace{2pt}} c@{\hspace{2pt}} c@{\hspace{2pt}} 
c@{\hspace{2pt}} c@{\hspace{2pt}} c@{\hspace{2pt}} c } 
  
  \tableline
   \noalign{\smallskip}
Region & Branch &  \Te & EP84 &  MRS85 & M91 & ZKH94 & P00 & P01$^a$&PT05$^b$& KD02 & KK04 & D02 & PP04a & PP04b & PP04c & N06a$^c$& N06b & N06c \\ 

& & &  R$_{23}$ & R$_{23}$ & R$_{23}$, $y$ & R$_{23}$ & R$_{23}$ & R$_{23}$, P & R$_{23}$, P & R$_{23}$, $q$ & R$_{23}$, $q$ & 
N$_2$ & N$_2$ & N$_2$ & O$_3$N$_2$ & R$_{23}$ & N$_2$ & O$_3$N$_2$  \\

\tableline
\noalign{\smallskip}  
HCG 31 AC &H&8.22$\pm$0.05&8.25& 8.89  & 8.67 & 8.74  & 8.47 & 8.15 & 8.09 & 7.99 & 8.12 & 8.40 & 8.34 & 8.29 & 8.30 & 8.05  &8.16& 8.22\\ 
HCG 31 B  &H&8.14$\pm$0.08&8.14& 8.48  & 8.29 & 8.44  & 8.24 & 8.22 & 8.12 & 8.41 & 8.44 & 8.39 & 8.33 & 8.28 & 8.29 & 8.07  &8.16& 8.20\\ 
HCG 31 E  &H&8.13$\pm$0.09&8.17& 8.62  & 8.14 & 8.53  & 8.31 & 8.18 & 8.13 & 8.19 & 8.32 & 8.35 & 8.30 & 8.26 & 8.25 & 8.07  &8.11& 8.15\\ 
HCG 31 F1 &L&8.07$\pm$0.06&8.02& 8.30  & 8.13 &\nodata& 7.86 & 8.12 & 7.99 & 8.46 & 8.33 & 8.05 & 8.07 & 8.09 & 8.03 &\nodata&7.81& 7.67\\ 
HCG 31 F2 &L&8.03$\pm$0.10&7.90& 8.54  & 8.06 & 8.48  & 7.76 & 8.13 & 7.95 & 8.19 & 8.27 & 8.06 & 8.07 & 8.10 & 8.07 & 8.07  &7.83& 7.79\\ 
HCG 31 G  &H&8.15$\pm$0.08&8.13& 8.43  & 8.26 & 8.40  & 8.22 & 8.11 & 8.17 & 8.31 & 8.42 & 8.41 & 8.34 & 8.29 & 8.26 & 8.07  &8.17& 8.16\\ 
Mkn 1199 C&H&8.75$\pm$0.12&9.37& 9.26  & 9.00 & 9.18  & 9.19 & 8.71 & 8.54 & 9.14 & 9.14 & 8.92 & 8.74 & 8.90 & 8.81 & 9.18  &8.78& 8.94\\ 
Haro 15 A &H&8.10$\pm$0.06&8.08&\nodata& 8.14 &\nodata& 7.91 & 8.12 & 8.12 & 8.48 & 8.34 & 7.98 & 8.01 & 8.05 & 7.97 &\nodata&7.74& 7.38\\ 
Mkn 5 A1  &L&8.07$\pm$0.07&7.90& 8.54  & 8.04 & 8.48  & 7.76 & 8.13 & 8.13 & 8.19 & 8.26 & 8.18 & 8.17 & 8.16 & 8.12 & 8.07  &7.94& 7.89\\ 
Mkn 5 A2  &L&8.08$\pm$0.07&7.95& 8.43  & 8.14 & 8.41  & 7.81 & 7.92 & 8.17 & 8.18 & 8.33 & 8.16 & 8.15 & 8.15 & 8.11 & 8.07  &7.92& 7.87\\ 
IRAS 08208+2816&H&8.33$\pm$0.08&8.15&8.50&8.55& 8.46  & 8.25 & 8.42 & 8.35 & 8.35 & 8.25 & 8.47 & 8.39 & 8.34 & 8.23 & 8.35  &8.23& 8.11\\ 
POX 4    &L&8.03$\pm$0.04&8.15&\nodata& 8.20 & \nodata&7.97 & 7.92 & 8.06 & 8.48 & 8.40 & 7.78 & 7.86 & 7.91 & 7.87 &\nodata&7.53&\nodata\\ 
UM 420    &L&7.95$\pm$0.05&7.78& 8.73  & 7.98 & 8.61  & 7.66 & 7.85 & 7.86 & 8.02 & 8.16 & 8.20 & 8.39 & 8.33 & 8.28 & 7.57  &8.15& 8.14\\ 
SBS 0926+606A&L&7.94$\pm$0.08&7.88&8.57& 7.97 & 8.50  & 7.75 & 7.77 & 7.80 & 8.17 & 8.20 & 7.97 & 8.00 & 8.05 & 8.02 & 7.71  &7.73& 7.64\\ 
SBS 0948+532&L&8.03$\pm$0.05&8.01& 8.31& 8.06 &\nodata& 7.86 & 7.82 & 8.10 & 8.34 & 8.28 & 7.91 & 7.95 & 8.01 & 7.95 &\nodata&8.01&\nodata\\ 
SBS 1054+365&L&8.00$\pm$0.07&8.05& 8.21& 8.09 &\nodata& 7.89 & 7.84 & 7.91 & 8.48 & 8.30 & 7.87 & 7.93 & 7.98 & 7.93 &\nodata&7.63&\nodata\\ 
SBS 1211+540&L&7.65$\pm$0.04&7.86& 8.60& 7.85 & 8.52  & 7.73 & 7.68 & 7.65 & 8.02 & 8.10 & 7.58 & 7.70 & 7.69 & 7.84 & 7.68  &7.31&\nodata\\ 
SBS 1319+579A&L&8.05$\pm$0.06&8.09&\nodata&8.13&\nodata&7.93 & 8.11 & 8.11 & 8.48 & 8.33 & 7.77 & 7.85 & 7.90 & 7.88 &\nodata&7.52&\nodata\\ 
SBS 1319+579B&L&8.12$\pm$0.10&7.85&8.62& 8.01 & 8.53  & 7.72 & 8.13 & 8.12 & 8.13 & 8.23 & 8.14 & 8.14 & 8.14 & 8.11 & 8.07  &7.90& 7.89\\ 
SBS 1319+579 C&L&8.15$\pm$0.07&7.85&8.62&8.02 & 8.53  & 7.72 & 8.13 & 8.13 & 8.12 & 8.23 & 8.18 & 8.17 & 8.16 & 8.13 & 8.06  &7.94& 7.93\\ 
SBS 1415+437 C&L&7.58$\pm$0.05&7.63&8.91&7.72 & 8.76  & 7.53 & 7.57 & 7.55 & 7.86 & 7.99 & 7.79 & 7.86 & 7.92 & 7.99 & 7.39  &7.55& 7.55\\ 
SBS 1415+437 A&L&7.61$\pm$0.06&7.58&8.96&7.64 & 8.80  & 7.49 & 7.50 & 7.48 & 7.82 & 7.92 & 7.72 & 7.81 & 7.86 & 7.97 & 7.34  &7.48& 7.41\\ 
III Zw 107 &H&8.23$\pm$0.09&8.17  &8.62&8.57 & 8.53   & 8.31 & 8.40 & 8.35 & 8.13 & 8.24 & 8.39 & 8.33 & 8.28 & 8.23 & 8.46  &8.15& 8.10\\ 
Tol 9 INT  &H&8.58$\pm$0.15&8.30& 9.00 & 8.76 & 8.84  & 8.58 & 8.61 & 8.55 & 8.95 & 8.90 & 8.68 & 8.56 & 8.54 & 8.42 & 8.77  &8.46& 8.40\\ 
Tol 9 NOT  &H&8.55$\pm$0.16&8.29& 8.97 & 8.73 & 8.81  & 8.55 & 8.56 & 8.50 & 8.94 & 8.88 & 8.72 & 8.59 & 8.59 & 8.44 & 8.75  &8.51& 8.44\\ 
Tol 1457-262A&L&8.05$\pm$0.07&8.09&\nodata&8.26&\nodata&7.92 & 8.11 & 8.20 & 8.58 & 8.42 & 8.04 & 8.06 & 8.09 & 8.02 &\nodata&7.80& 7.63\\ 
Tol 1457-262B&L&7.88$\pm$0.07&8.55&\nodata&8.21&\nodata&7.87 & 7.91 & 8.21 & 8.38 & 8.29 & 7.88 & 7.93 & 7.99 & 7.96 &\nodata&7.60&\nodata\\ 
Tol 1457-262C&L&8.06$\pm$0.11&8.48&\nodata&8.21&\nodata&7.88 & 8.00 & 8.24 & 8.48 & 8.37 & 8.07 & 8.08 & 8.10 & 8.06 &\nodata&7.83& 7.77\\ 
ESO 566-8 &H&8.46$\pm$0.11&8.27& 8.92 & 8.68 &  8.77  & 8.50 & 8.44 & 8.38 & 8.92 & 8.84 & 8.84 & 8.68 & 8.76 & 8.51 & 8.70  &8.66& 8.53\\ 
NGC 5253 A&H&8.18$\pm$0.04&8.09& 8.24 & 8.13 &\nodata & 8.15 & 8.11 & 8.13 & 8.53 & 8.33 & 8.40 & 8.34 & 8.28 & 8.17 &\nodata&8.16& 8.00\\ 
NGC 5253 B&H&8.19$\pm$0.04&8.09& 8.21 & 8.14 &\nodata & 8.14 & 8.11 & 8.13 & 8.48 & 8.34 & 8.34 & 8.29 & 8.25 & 8.14 &\nodata&8.10& 7.94\\ 
NGC 5253 C&L&8.28$\pm$0.04&8.14& 8.46  & 8.53 & 8.42  & 8.23 & 8.38 & 8.32 & 8.67 & 8.63 & 8.11 & 8.11 & 8.13 & 8.08 & 8.30  &7.87& 7.80\\ 
NGC 5253 D&L&8.31$\pm$0.07&8.15& 8.52  & 8.19 & 8.47  & 8.26 & 8.23 & 8.17 & 8.32 & 8.37 & 8.31 & 8.27 & 8.23 & 8.22 & 8.37  &8.07& 8.10\\ 
\tableline
  \end{tabular}
    \begin{flushleft}
 NOTE: The empirical calibrations and the parameters used for each of them are:   
 EP84: Edmunds \& Pagel (1984) that involves the $R_{23}$ parameter; MRS85: McCall, Rybski \& Shields (1985) using
$R_{23}$; M91: McGaugh (1991) using $R_{23}$ and $y$; ZKH94: Zaritzky, Kennicutt \& Huchra (1994) using $R_{23}$; P00: Pilyugin (2000) using 
$R_{23}$; P01: Pilyugin (2001a,b) using $R_{23}$ and $P$; KD02: Kewley \& Dopita (2002) using $R_{23}$ and $q$; KK04: Kobulnicky \& Kewley (2004) 
using $R_{23}$ \& $q$; D02: Denicol\'o, Terlevich \& Terlevich (2002) using the $N_2$ parameter; PP04: Pettini \& Pagel (2004), using (a) $N_2$ with 
a linear fit, (b) $N_2$ with a cubic fit, (c) the $O_3N_2$ parameter; N06: Nagao et al. (2006) using the cubic relations involving the $R_{23}$ (a), 
$N_2$ (b) and $O_3N_2$ (c) parameters. The value compiled in the column labeled \Te\ is the oxygen abundance derived by the direct method. \\

  $^a$ The value listed for P01 is the average value between the high- and the low-metallicity branches for objects with 7.90$<$\abox$<$8.20.\\
  $^b$ The value listed for PT05 is the average value between the high- and the low-metallicity branches for objects with 8.05$<$\abox$<$8.20.\\
  $^c$ The value listed for N06 is the average value between the high- and the low-metallicity branches for objects with 8.00$<$\abox$<$8.15.\\
  \end{flushleft}
  \end{table*}

\section{Comparison with empirical calibrations}


\begin{figure*}[t!]
\centering
\begin{tabular}{cc}
\includegraphics[angle=270,width=0.4\linewidth]{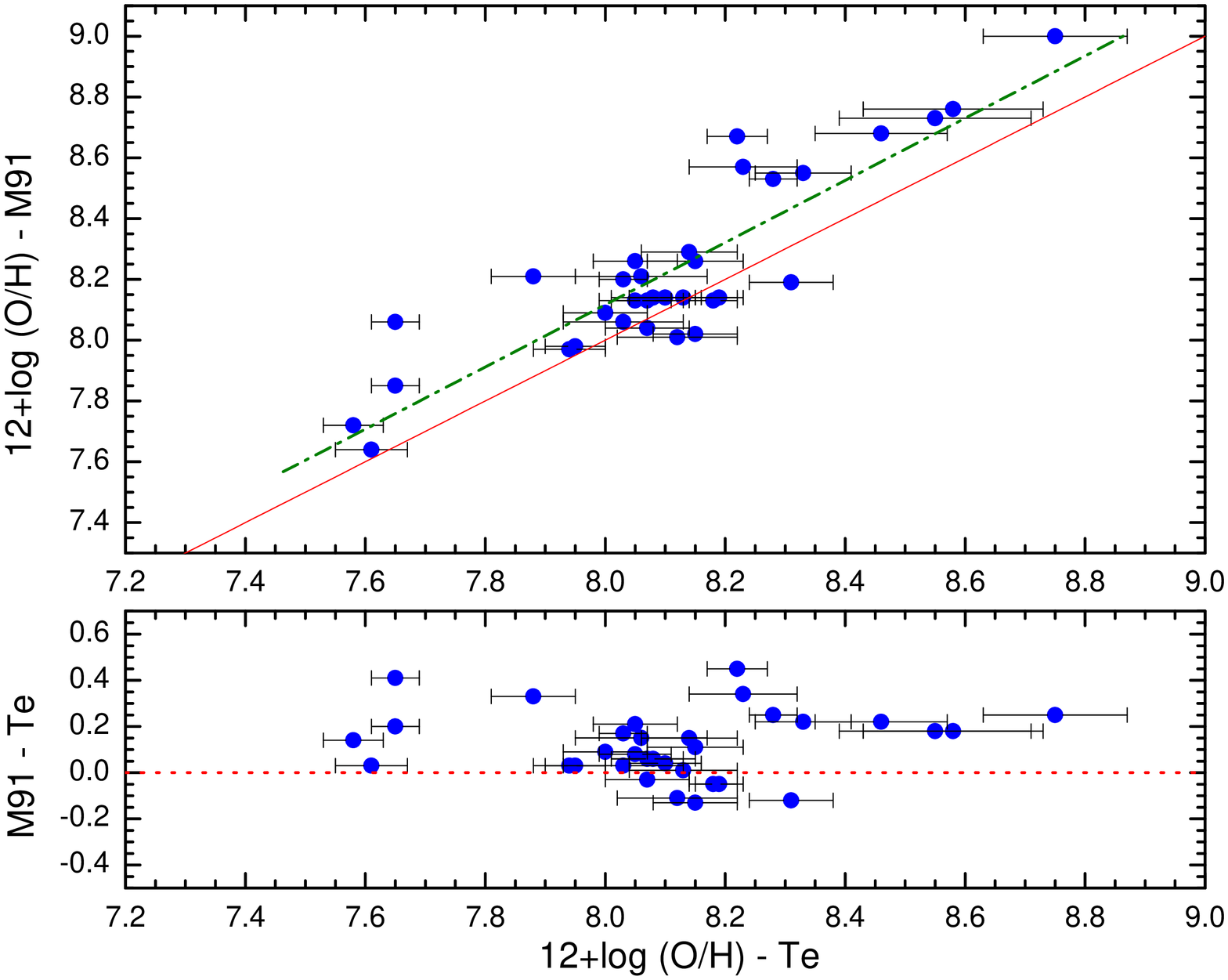} &  
\includegraphics[angle=270,width=0.4\linewidth]{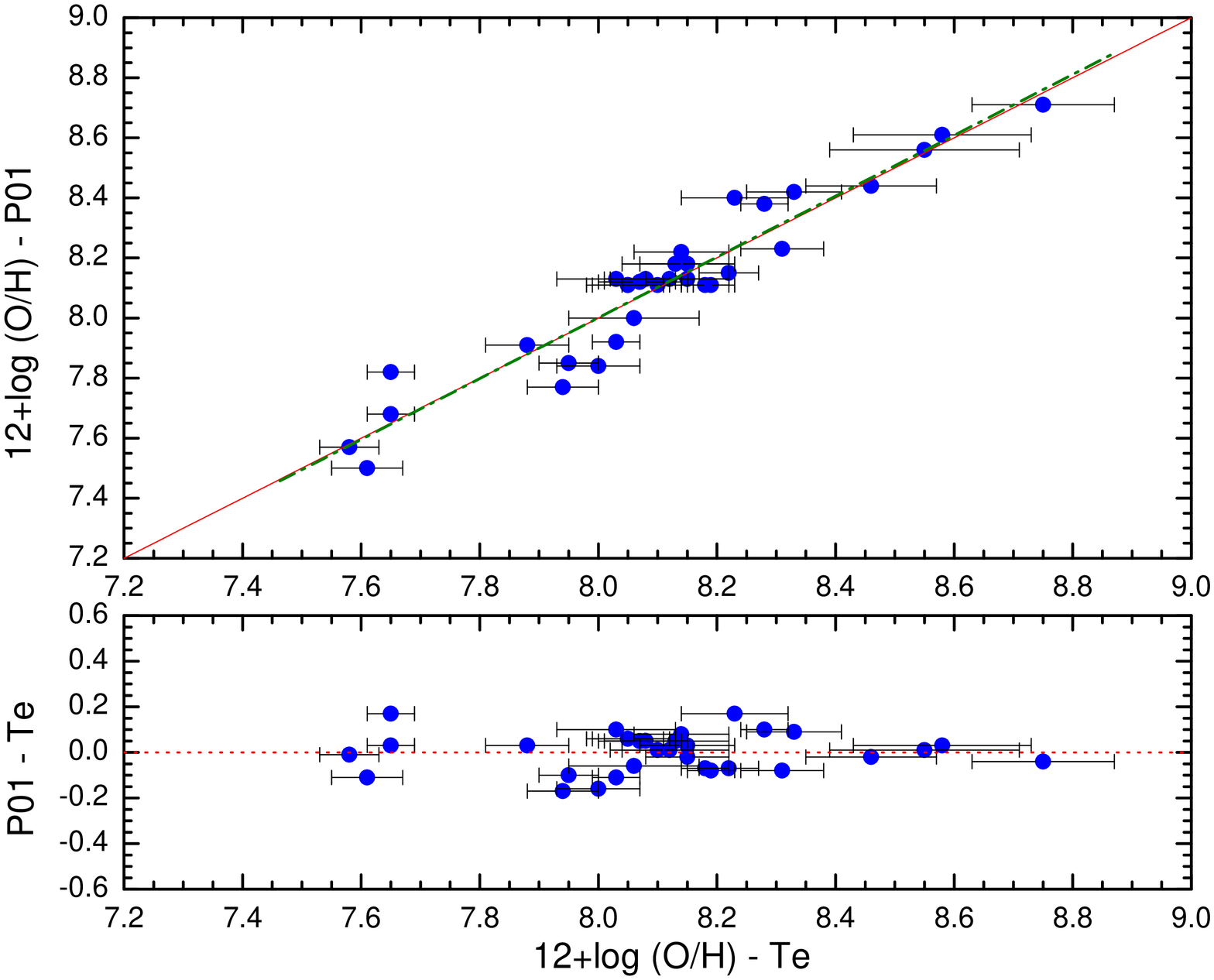} \\  
\includegraphics[angle=270,width=0.4\linewidth]{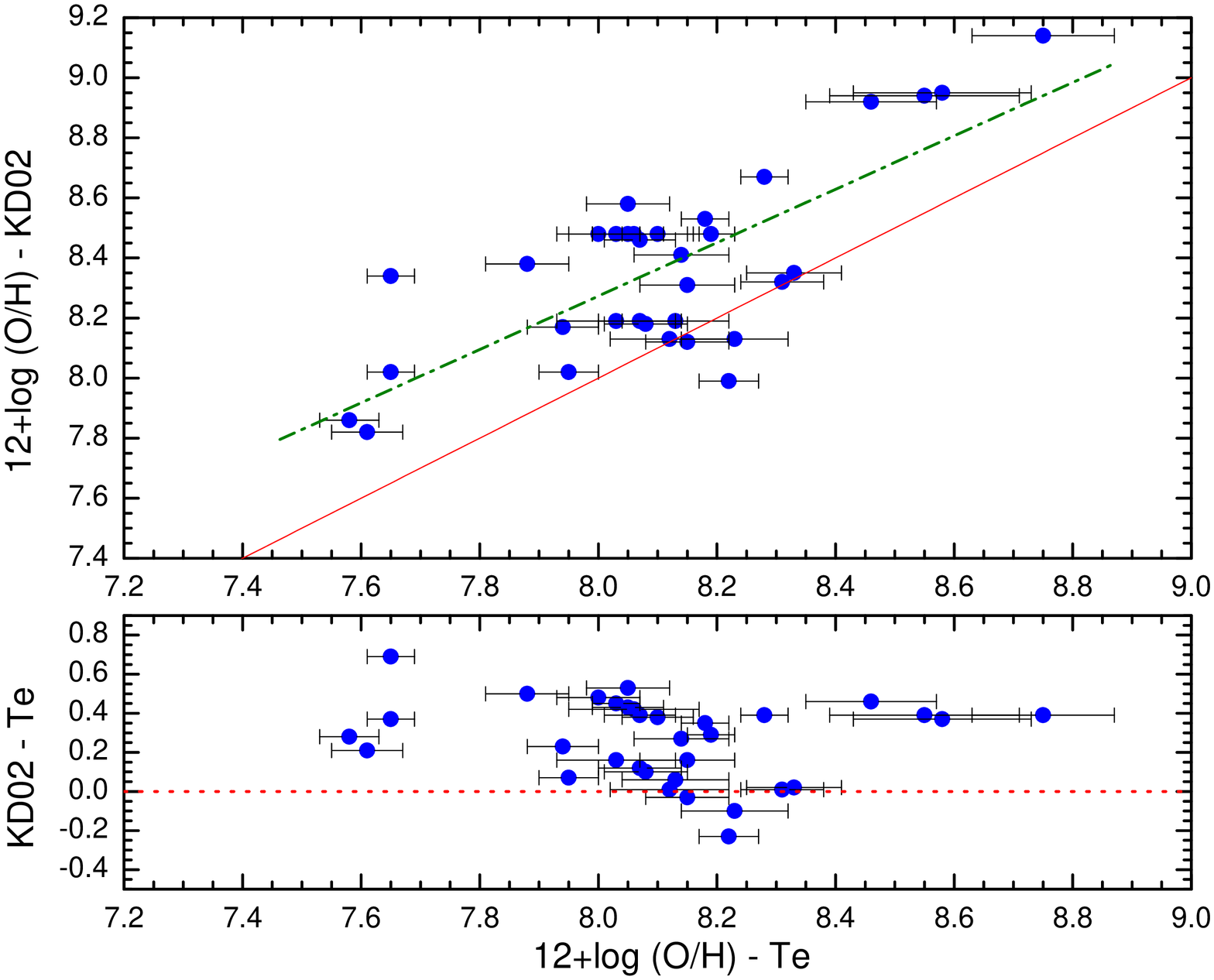} &  
\includegraphics[angle=270,width=0.4\linewidth]{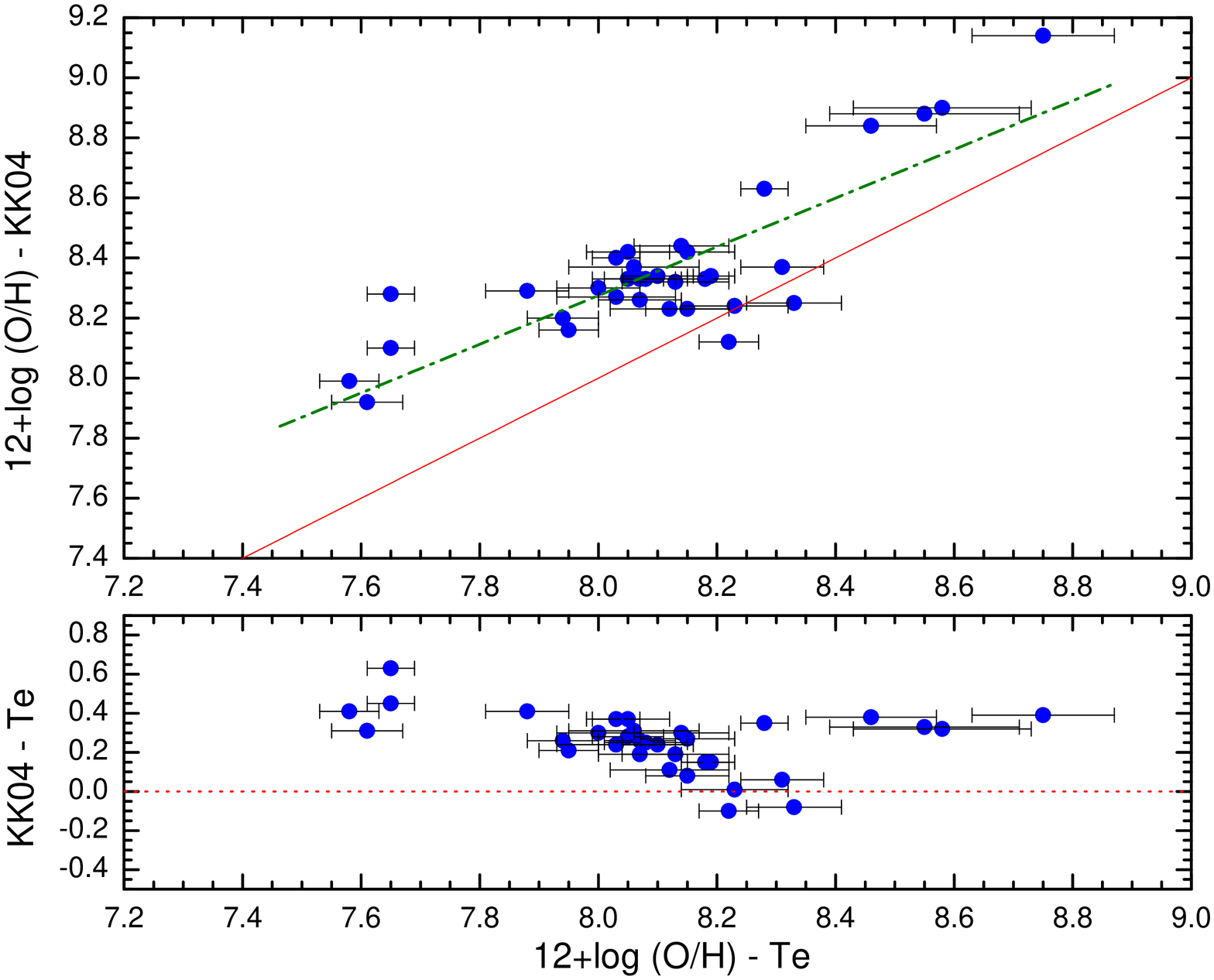} \\  
\includegraphics[angle=270,width=0.4\linewidth]{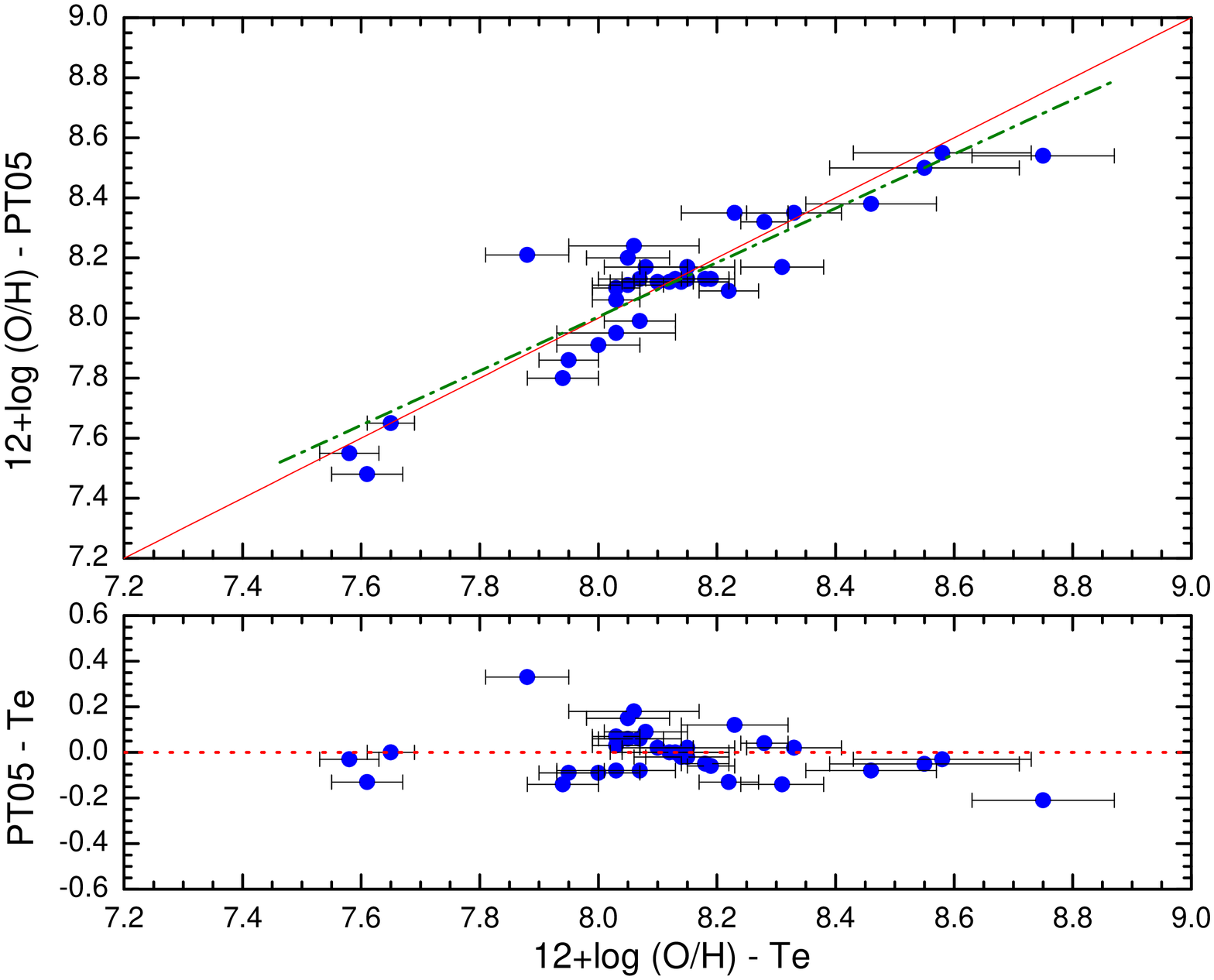} &  
\includegraphics[angle=270,width=0.4\linewidth]{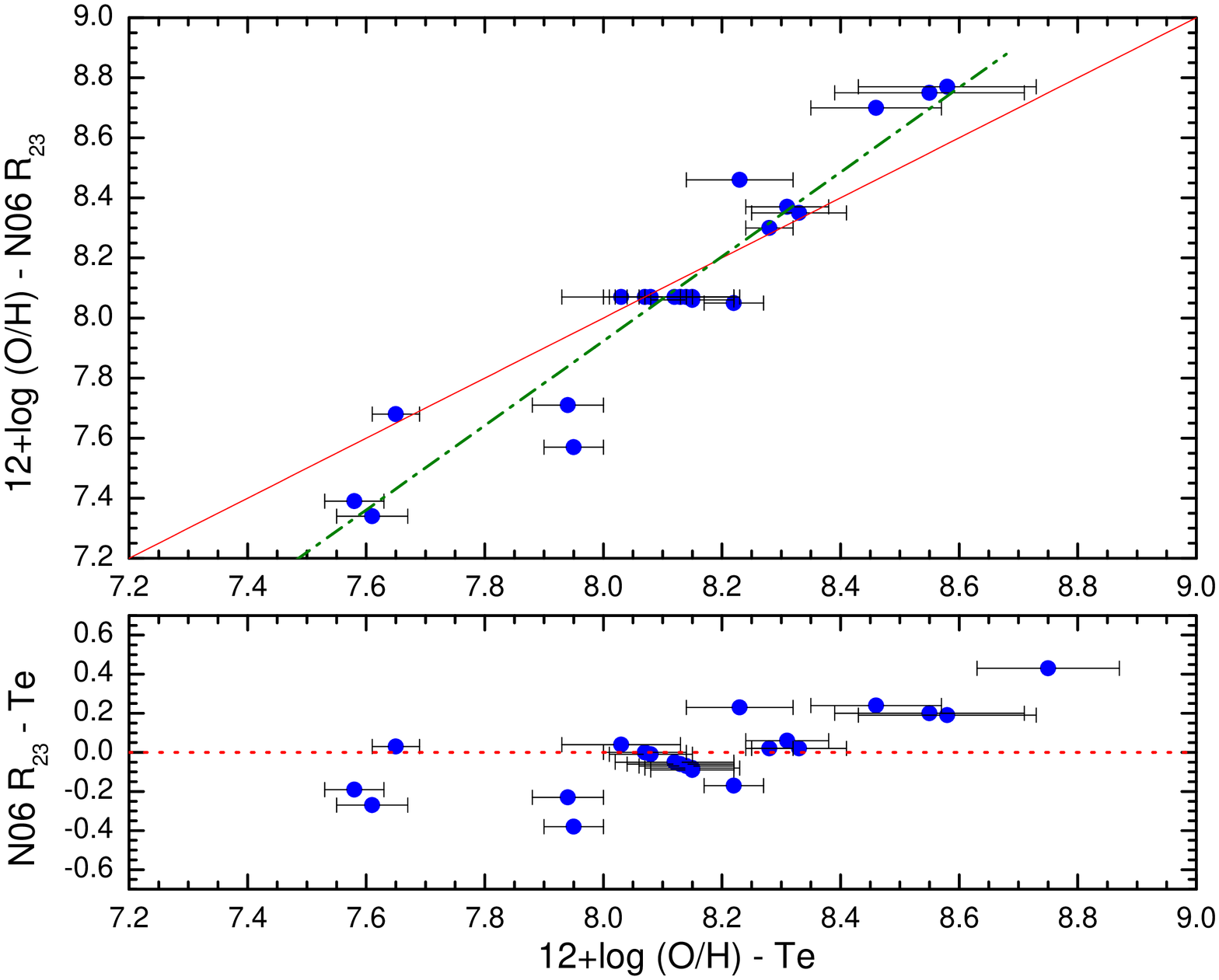} \\  
\end{tabular}
\protect\caption[ ]{\footnotesize{Comparison between the oxygen abundances derived using the direct method (\Te, always plotted in the $x-axis$) with 
those estimated using six different empirical calibrations that consider the $R_{23}$ parameter: M91: \citet{McGaugh91}; P01: \citet{P01a,P01b}; 
KD02: \citet{KD02}; KK04: \citet{KK04}; PT05: \citet{PT05}; N06: \citet*{Nagao06}. The bottom panel of each diagram indicates the difference between 
empirical and direct data.}}
\label{figemp}
\end{figure*}

We used the data of the 31 regions for which we have a direct estimate of \Te\ and, hence, a direct estimate of the oxygen abundance --see \citet{LSE10b} and their Table~3-- , to 
check the reliability of several empirical calibrations.
A recent review of 10 metallicity calibrations, including theoretical and empirical methods, was presented by \citet{KE08}, but
previous section
gives an overview of the most common empirical calibrations and defines the typical parameters that are used to estimate the oxygen 
abundance following these relations. These parameters are ratios between bright emission lines, the most commonly used are $R_{23}$, $P$, $y$, $N_2$, 
and $O_3N_2$ 
Table~\ref{abempirica1} lists the values of all these parameters derived for each region with a direct estimate of the oxygen abundance --see \citet{LSE09} 
 for details--. Table~\ref{abempirica1} also includes the value derived for the $q$ parameter (in units of cm s$^{-1}$) obtained from the 
optimal calibration provided by \citet{KD02}. The results for the oxygen abundances derived for each object and empirical calibration are listed in 
Table~\ref{abempirica2}. This table also indicates the branch (high or low metallicity) considered in each region when using the $R_{23}$ parameter
although, as is clearly specified in the table, for some objects with 8.00$\leq$\abox$\leq$8.3 we assumed the average value
found for the lower and upper branches.

Looking at the data compiled in Table~\ref{abempirica2} the huge range of oxygen abundance found for the same object using different 
calibrations is evident. As \citet{KE08} concluded, it is critical to use the same metallicity calibration when comparing properties from different data sets or 
investigate luminosity-metallicity or mass-metallicity relations. 
Furthermore, abundances derived with such strong-line methods may be significantly biased if the objects under study have different structural 
properties (hardness of the ionizing radiation field, morphology of the nebulae) than those used to calibrate the methods \citep{Stasinska09}.

\begin{figure*}[t!]
\centering
\begin{tabular}{cc}
\includegraphics[angle=270,width=0.4\linewidth]{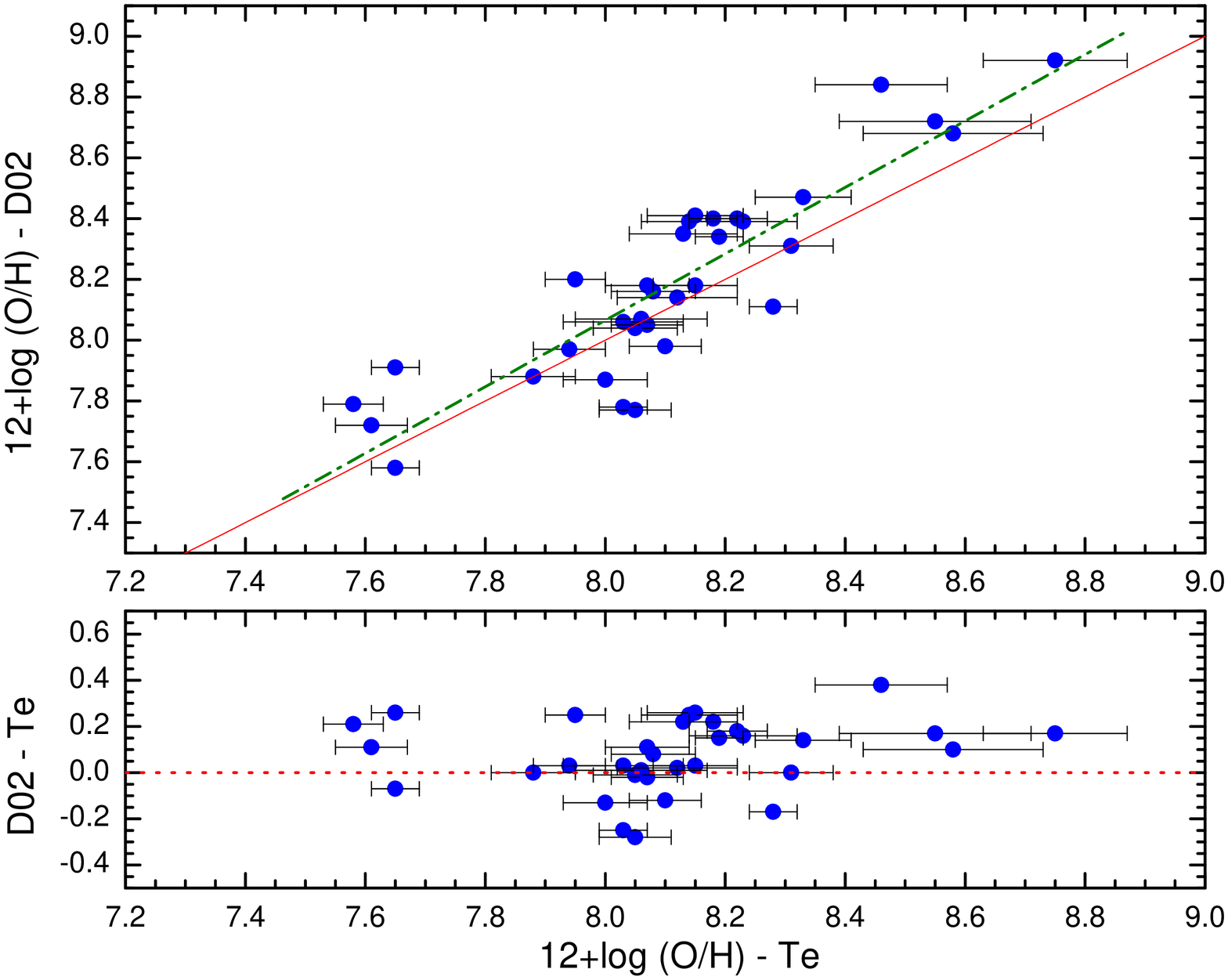} &  
\includegraphics[angle=270,width=0.4\linewidth]{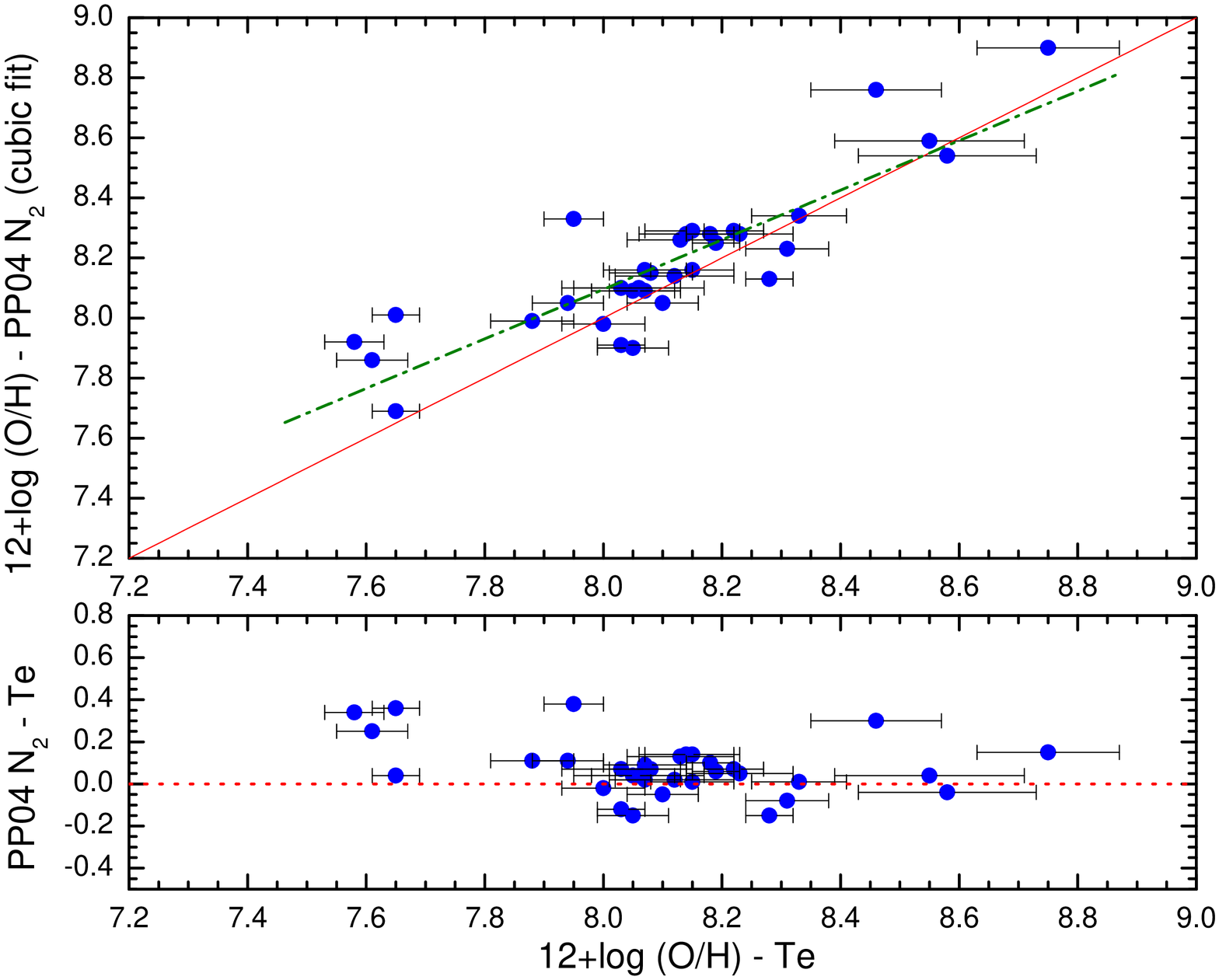} \\  
\includegraphics[angle=270,width=0.4\linewidth]{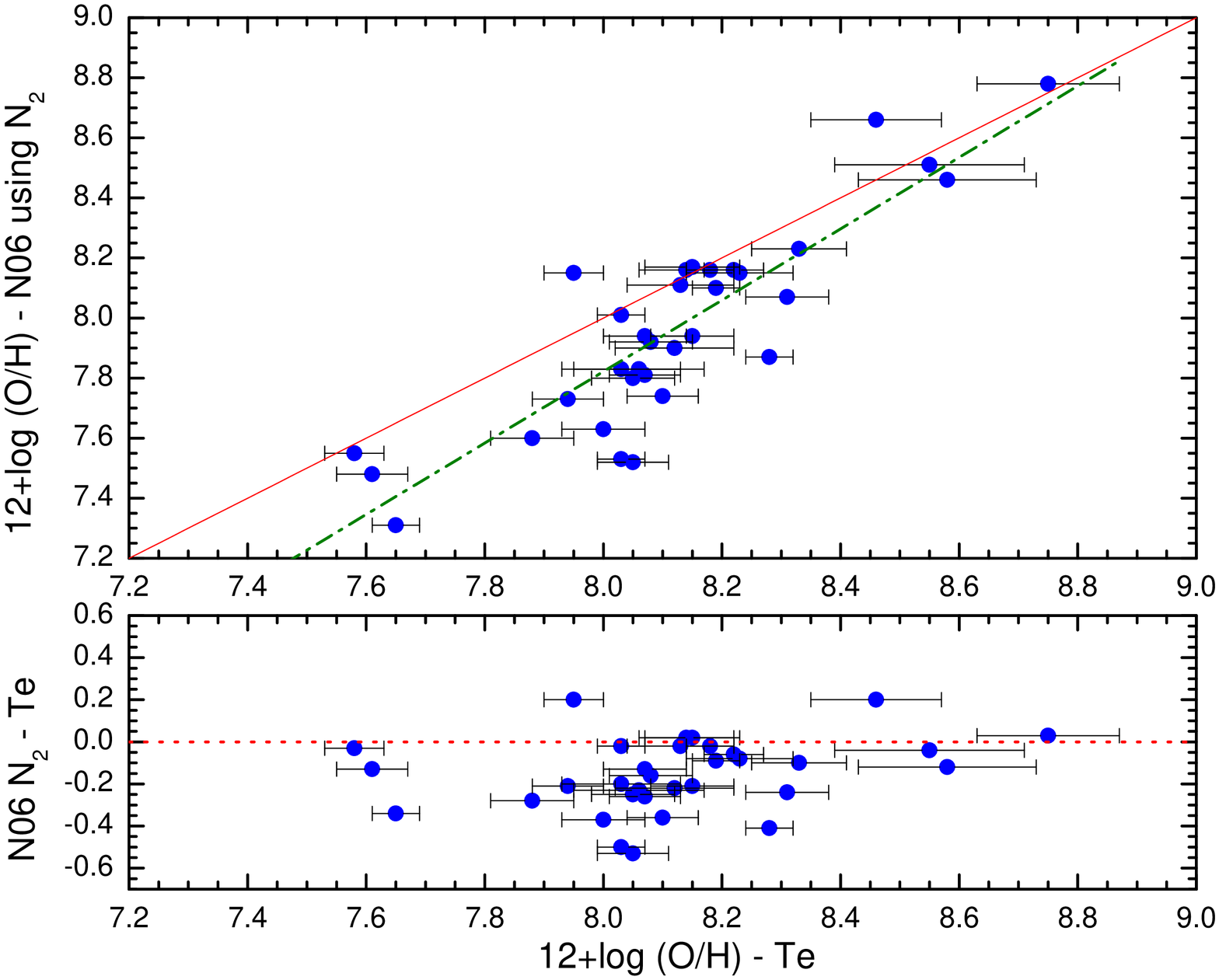} &  
\includegraphics[angle=270,width=0.4\linewidth]{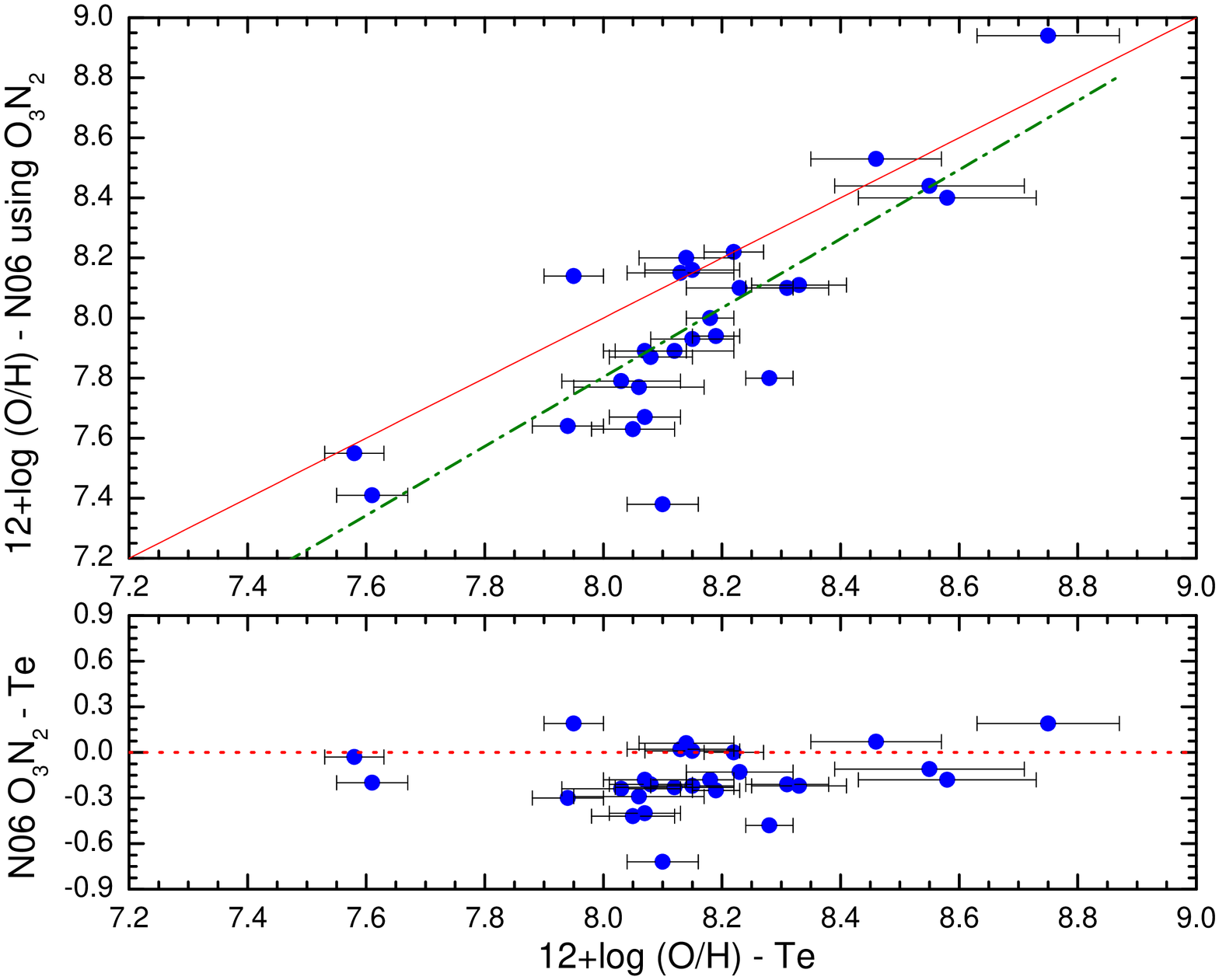} \\  
\end{tabular}
\protect\caption[ ]{\footnotesize{Comparison between the oxygen abundances derived using the direct method (\Te, in the $x-axis$) with those computed 
using the calibrations using the $N_2$ parameter --D02: \citet{D02}; PP04b: \citet{PP04} considering their cubic fit to $N_2$; N06: 
\citet*{Nagao06}-- and the O$_3$N$_2$ parameter following the cubic fit provided by \citet*{Nagao06}. The bottom panel of each diagram indicates the 
difference between empirical and direct data.}}
\label{figemp2}
\end{figure*}

Figures~\ref{figemp} and \ref{figemp2} plots the ten most common calibrations and their comparison with the oxygen abundance obtained using the 
direct method.
We performed a simple statistic analysis of the results to quantify the goodness of these empirical calibrations. Table~\ref{dispempi} compiles 
the average value and the dispersion (in absolute values) of the difference between the abundance given by empirical calibration and that obtained 
using the direct method. We check that the empirical calibration that provides the best results is that proposed by \citet{P01a,P01b}, which gives 
oxygen abundances very close to the direct values (the differences are lower than 0.1 dex in the majority of the objects), and furthermore it 
possesses a low dispersion. 
We note however that the largest divergences found using this calibration are in the low-metallicity regime. The update of this calibration 
presented by \citet{PT05} seems to partially solve this problem, the abundances provided by this calibration also agree very well with those derived 
following the direct method. We therefore conclude that the \citet{PT05} calibration is nowadays the best suitable method to derive the oxygen 
abundance of star-forming galaxies when auroral lines are not observed.    

On the other hand, the results given by the empirical calibrations provided by \citet{McGaugh91}, \citet{KD02} and \citet{KK04}, that are based on 
photoionization models, are systematically higher than the values derived from the direct method. This effect is even more marked in the last two 
calibrations, which usually are between 0.2 and 0.3 dex higher than the expected values. These empirical calibrations also have a higher dispersion 
than that estimated for \citet{P01a,P01b} or \citet{PT05} calibrations. \citet{Yin+07} also found high discrepancies when comparing the theoretical 
metallicities using the theoretical models of \citet{Tremonti04} with the \Te-based metallicites obtained from \citet{P01a,P01b} and \citet{PT05}.

One of the possible explanations for the different metallicities obtained between the direct method and those derived from the empirical calibrations 
based on photoionization models are temperature fluctuations in the ionized gas. Temperature gradients or fluctuations indeed cause 
the true metallicities based on the \Te-method to be underestimated (i.e. Peimbert 1967; Stasinska 2002,2005; Peimbert et al. 2007). 
Temperature fluctuations can also explain our results for NGC~5253 \citep{LSEGRPR07}: the ionic abundances of O$^{++}$/H$^+$ and 
C$^{++}$/H$^+$ derived from recombination lines are systematically 0.2 -- 0.3 dex higher than those determined from the direct method --based on the 
intensity ratios of collisionally excited lines. This abundance discrepancy has been also found in Galactic \citep{GRE07} and other extragalactic 
\citep{Esteban09} \HII regions and interestingly this discrepancy is in all cases of the same order as the differences between abundances 
derived from the direct methods and empirical calibrations based on photoionization models.

The conclusion that temperature fluctuations do exist in the ionized gas of starburst galaxies is very important for the analysis of the chemical 
evolution of galaxies and the Universe. Indeed, if that is correct, the majority of the abundance determinations in extragalactic objects following the 
direct method, including those provided in this work, have been underestimated by at least 0.2 to 0.3 dex. Deeper observations of a large 
sample of star-forming galaxies --that allow us to detect the faint recombination lines, such as those provided by \citet{Esteban09}-- and more 
theoretical work --including a better understanding of the photoionization models, such as the analysis provided by \citet{KE08}-- are needed to confirm 
this puzzling result.

\begin{table*}[t!]
\centering
  \caption{\footnotesize{Results of the comparison between the oxygen abundance given by several empirical calibrations and the oxygen abundance derived here following the direct (\Te) method.}}
  \label{dispempi}
  \begin{tabular}{l@{\hspace{8pt}}  c@{\hspace{8pt}}c@{\hspace{8pt}}c@{\hspace{8pt}} c@{\hspace{8pt}}c@{\hspace{8pt}} c@{\hspace{8pt}}  c 
c@{\hspace{8pt}}  c@{\hspace{8pt}} c@{\hspace{8pt}} c c@{\hspace{8pt}} c@{\hspace{8pt}} } 
  \tableline
   \noalign{\smallskip}
Parameter   & \multicolumn{6}{c@{\hspace{4pt}}}{$R_{23}$} & & \multicolumn{3}{c@{\hspace{4pt}}}{$N_2$} & & 
\multicolumn{2}{c@{\hspace{4pt}}}{$O_3N_2$} \\

   \cline{2-7} \cline{9-11} \cline{13-14}
         \noalign{\smallskip}
		 
Calibration$^a$ &  P01   & PT05  &  N06  & M91  & KD02 & KK04  & & D02  &  PP04 &   N06 & & PP04  &  N06    \\ 

\tableline
\noalign{\smallskip}    
	Average$^b$ &  0.07  & 0.08 &  0.14  & 0.15 & 0.28 & 0.27  & & 0.14    & 0.12    &  0.18 & & 0.12 & 0.21    \\ 
 $\sigma^c$   &  0.05  & 0.07 &  0.12  & 0.11 & 0.18 & 0.13  & & 0.10    & 0.10    &  0.14 & & 0.10 & 0.16   \\ 
 Notes$^d$      & B/A (1)& B/A  &   (2)  & S.H. & S.H. & S.H.  & & S.H. (3)& B/A (4) &  S.L.(5)  & & B/A (6) & S.L.      \\ 
\noalign{\smallskip}
\tableline
  \end{tabular}
  \begin{flushleft}
  $^a$ The names of the calibrations are the same as in Table~\ref{abempirica2}.\\
  $^b$ Average value (in absolute values) of the difference between the abundance given by empirical calibrations 
and that obtained using the direct method. The names of the calibrations are the same as in Table~\ref{abempirica2}. \\
 $^c$ Dispersion (in absolute values) of the difference between the abundance given by empirical calibrations and that obtained using the direct method.\\
 $d$ We indicate if the empirical calibration gives results both below and above the direct value (B/A), if they are systematically higher than the direct value (S.H.) or if they are systematically lower than the direct value (S.L.). Some additional notes are: \\ 
  (1) Higher deviation in the low branch. \\ 
  (2) This calibration provides lower oxygen abundances in low-metallicity regions and higher oxygen abundances in high-metallicity regions.\\ 
  (3) Systematically higher only for \abox$>$8.2. \\ 
  (4) Higher deviation for  \abox$<$8.0. Considering \abox$>$8.0, we get average=0.08 and $\sigma$=0.06. \\ 
  (5) Higher deviation at lower oxygen abundances. \\
  (6) Higher deviation for \abox$<$8.0. Assuming \abox$>$8.0, average=0.09 and $\sigma$=0.06. 
  \end{flushleft}
\end{table*}

On the other hand, we have checked the validity of the recent relation provided by \citet*{Nagao06}, which merely considers a cubic fit between the 
$R_{23}$ parameter and the oxygen abundance. This calibration was obtained combining data from several large galaxy samples, the majority from the 
SDSS, which includes all kinds of star-forming objects. As it is clearly seen in Table~\ref{abempirica2} and in Fig.~\ref{figemp2}, the 
\citet*{Nagao06} relation is not suitable to derive a proper estimate of the oxygen abundance for the majority of the objects in our galaxy sample. 
In general, this calibration provides lower oxygen abundances in low-metallicity regions and higher oxygen abundances in high-metallicity regions. 
Objects located in the metallicity range 8.00$\leq$\abox$\leq$8.15 have systematically 12+log(O/H)$_{\rm N06}\sim$8.07 because we have to use an 
average value between the low and the high branches. Furthermore, many of the regions do not have a formal solution to the \citet*{Nagao06} equation, 
such as the maximum value for $R_{23}$ is 8.39 at \abox=8.07. We consider that the use of an ionization parameter --$P$ as introduced by 
\citet{P01a,P01b} or $q$ as followed by \citet{KD02}-- is fundamental to obtain a real estimate of the oxygen abundance in star-forming galaxies, 
especially in objects showing strong starbursts. In the same sense, the direct method and not the formulae provided by \citet{Izotov06} (which assumes 
a low-density approximation in order not to have to solve the statistical equilibrium equations of the O$^{+2}$ ion) provides a good approximation to 
the actual oxygen abundance when the auroral line [\ion{O}{iii}] $\lambda$4653 is observed.

Empirical calibrations considering a linear fit to the $N_2$ ratio \citep{D02,PP04} give results that are systematically $\sim$0.15 dex higher that 
the oxygen abundances derived from the direct method. The difference is higher at higher metallicities. 
We do not consider that this trend is a consequence of comparing different objects: both \citet*{D02} and \citet{PP04} calibrations are 
obtained using a sample of star-forming galaxies similar to those analysed in this work, many of which are WR galaxies. \citet*{D02} compared the 
$N_2$ ratio with the ionization parameter together with the results of photoionization models and concluded that most of the observed trend of $N_2$ 
with the oxygen abundance is caused by metallicity changes.
The cubic fit to $N_2$ performed by \citet{PP04} better reproduces the oxygen abundance, especially in the intermediate- and high-metallicity regime  
(\abox$>$8.0), where it has an average error of $\sim$0.08 dex. However, the cubic fit to $N_2$ provided by \citet*{Nagao06} gives systematically 
lower values for the oxygen abundance than those derived using the direct method, having an average error of $\sim$0.18 dex. 

The empirical calibration between the oxygen abundance and the $O_3N_2$ parameter proposed by \citet{PP04} gives acceptable results for objects with 
\abox$>$8.0, with the average error $\sim$0.1 dex. However, the new relation provided by \citet*{Nagao06} involving the $O_3N_2$  parameter gives 
systematically lower values for the oxygen abundance. As we commented before, we consider that the \citet*{Nagao06} calibrations are not suitable for 
studying galaxies with strong star-formation bursts. Their procedures must be taken with caution, galaxies with different ionization parameters, 
different chemical evolution histories, and different star formation histories should have different relations between the bright emission lines and 
the oxygen abundance. This issue is even more important when estimating the metallicities of intermediate- and high-redshift galaxies, because the 
majority of their properties are highly unknown.

\section{Conclusions}

 We compared the abundances provided by the direct method with those obtained using the most common empirical calibrations in our sample of star-forming regions within Wolf-Rayet galaxies --see \citet{LSE10b}--. The main conclusions are:
   \begin{itemize}
   \item The Pilyugin-method of \citet{P01a,P01b}, which considers the $R_{23}$ and the $P$ parameters and is updated by \citet{PT05}, is nowadays the best 
suitable empirical calibration to derive the oxygen abundance of star-forming galaxies. The cubic fit to $R_{23}$ provided by \citet*{Nagao06} is not 
valid for analysing these star-forming galaxies. 
   \item The relations between the oxygen abundance and the $N_2$ or the $O_3N_2$ parameters provided by \citet{PP04} give acceptable results for 
objects with \abox$>$8.0. 
   \item The results provided by empirical calibrations based on photoionization models \citep{McGaugh91,KD02,KK04} are systematically 0.2 -- 0.3 dex 
higher than the values derived from the direct method. These differences are of the same order as the abundance discrepancy 
found between abundances determined from recombination and collisionally excited lines of heavy-element ions.
This may suggest temperature fluctuations in the ionized gas, as they exist in Galactic and other extragalactic \HII\ regions.
   \end{itemize}

\twocolumn

\begin{acknowledgements}

\'A.R. L-S thanks C.E. (his formal PhD supervisor) for the help and very valuable explanations, talks and discussions during these years. 
He also acknow\-ledges the help and support given by the Instituto de Astrof\'{\i}sica de Canarias (Spain) while doing his PhD.
\'A.R. L-S. \emph{deeply} thanks the Universidad de La Laguna (Tenerife, Spain) for force him to translate his PhD thesis from English to Spanish; he 
had to translate it from Spanish to English to complete this publication. 
This was the main reason of the delay of the publication of this research, because the main results shown here were already included in the PhD 
dissertation (in Spanish) which the first author finished in 2006 \citep{LS06}. \'A.R. L-S. also thanks the people at the CSIRO/Australia 
Telescope National Facility, especially B\"arbel Koribalski, for their support and friendship while translating his PhD. 
This work has been partially funded by the Spanish Ministerio de Ciencia y Tecnolog\'{\i}a (MCyT) under project AYA2004-07466. 
This research has made use of the NASA/IPAC Extragalactic Database (NED) which is operated by the Jet Propulsion Laboratory, California Institute of 
Technology, under contract with the National Aeronautics and Space Administration.
This research has made extensive use of the SAO/NASA Astrophysics Data System Bibliographic Services (ADS).

\end{acknowledgements}


\begin{thebibliography}{}

\scriptsize{

\bibitem[\protect\citeauthoryear{Alloin et al.}{Alloin et al.}{1979}]{Alloin79} 
Alloin D., Collin-Souffrin S., Joly M. \& Vigroux L. 1979, A\&A, 78, 200 

\bibitem[\protect\citeauthoryear{Crowther}{Crowther}{2007}]{Crowther07}
Crowther, P.A. 2007, ARAA, 45, 177


\bibitem[\protect\citeauthoryear{Denicol\'o et al.}{Denicol\'o, Terlevich \& Terlevich}{2002}]{D02}
Denicol\'o, G., Terlevich, R. \& Terlevich, E. 2002, \mnras, 330, 69
\bibitem[\protect\citeauthoryear{de Naray et al.}{de Naray, McGaugh \& de Blok}{2004}]{deNaray04}
de Naray R.~K., McGaugh S.~S. \& de Blok  W.~J.~G., 2004, MNRAS, 355, 887 
\bibitem[\protect\citeauthoryear{De Rossi, Tissera \& Scannapieco}{De Rossi et al.}{2006}]{DeRossi06}
De Rossi, M.E., Tissera, P.B., \& Scannapieco, C. 2006, MNRAS, 374, 323
\bibitem[\protect\citeauthoryear{D\'{\i}az \& P\'erez-Montero}{D\'{\i}az \& P\'erez-Montero}{2000}]{Dia00}
D\'{\i}az, A.I., \& P\'erez-Montero, E. 2000, MNRAS, 312, 130


\bibitem[\protect\citeauthoryear{Dopita \& Evans}{Dopita \& Evans}{1986}]{DE86}
Dopita, M.A. \& Evans, I. N. 1986, ApJ, 307, 431
\bibitem[\protect\citeauthoryear{Dopita et al.}{Dopita et al.}{2000}]{Do00}
Dopita, M.A., Kewley, L. J., Heisler, C.A. \& Sutherland, R.S. 2000, ApJ, 542, 224

\bibitem[\protect\citeauthoryear{Edmunds \& Pagel}{Edmunds \& Pagel}{1978}]{EP78}
Edmunds, M.G. \& Pagel, B.E.J. 1978, MNRAS, 185, 77
\bibitem[\protect\citeauthoryear{Edmunds \& Pagel}{Edmunds \& Pagel}{1984}]{EP84}
Edmunds, M.G. \& Pagel, B.E.J. 1984, MNRAS, 211, 507


\bibitem[\protect\citeauthoryear{Erb et al.}{Erb et al.}{2006}]{Erb06}
Erb, D.K., Shapley, A.E., Pettini, M., Steidel, C.C., Reddy, N.A. \& Adelberger, K.L. 2006, ApJ, 644, 813
\bibitem[\protect\citeauthoryear{Esteban et al.}{Esteban et al.}{2009}]{Esteban09}
Esteban, C., Bresolin, F., Peimbert, M., Garc\'{\i}a-Rojas, J., Peimbert, A. \& Mesa-Delgado, A. 2009, ApJ, 700, 654

\bibitem[\protect\citeauthoryear{Ferland et al.}{Ferland et al.}{1998}]{Ferland98}
Ferland G.~J., Korista K.~T., Verner  D.~A., Ferguson J.~W., Kingdon J.~B. \& Verner E.~M, 1998, PASP, 110, 761 
\bibitem[\protect\citeauthoryear{Fioc \& Rocca-Volmerange}{Fioc \& Rocca-Volmerange}{1997}]{PEGASE97}
Fioc, M. \& Rocca-Volmerange, B. 1997, A\&A 326, 950

\bibitem[\protect\citeauthoryear{Garc\'{\i}a-Rojas \& Esteban}{Garc\'{\i}a-Rojas \& Esteban}{2007}]{GRE07}
Garc\'{\i}a-Rojas, J. \& Esteban, C., 2007, ApJ, 670, 457
\bibitem[\protect\citeauthoryear{Garnett}{Garnett}{1992}]{G92}
Garnett, D.R. 1992, AJ, 103, 1330

\bibitem[\protect\citeauthoryear{Hirashita et al.}{Hirashita et al.}{2001}]{Hirashita01}
Hirashita, H., Inoue, A.K., Kamaya, H. \& Shibai, H. 2001, A\&A, 366, 83


\bibitem[\protect\citeauthoryear{Izotov et al.}{Izotov et al.}{2006}]{Izotov06}
Izotov, Y.I., Stasi\'nska, G., Meynet, G., Guseva, N.G. \& Thuan, T.X. 2006, A\&A, 448, 955


\bibitem[\protect\citeauthoryear{Jensen et al.}{Jensen, Strom \& Strom}{1976}]{Jensen76}
Jensen E.~B., Strom K.~M. \& Strom S.~E. 1976, ApJ, 209, 748 

\bibitem[\protect\citeauthoryear{Kewley et al.}{Kewley et al.}{2001}]{KD01}
Kewley, L.J., Dopita, M.A., Sutherland, R.S., Heisler, C.A. \& Trevena, J. 2001, ApJS, 556, 121
\bibitem[\protect\citeauthoryear{Kewley \& Dopita}{Kewley \& Dopita}{2002}]{KD02}
Kewley, L.J. \& Dopita, M.A. 2002, ApJS, 142, 35
\bibitem[\protect\citeauthoryear{Kewley \& Ellison}{Kewley \& Ellison}{2008}]{KE08}
Kewley, L.J., \& Ellison, S.E. 2008, ApJ, 681, 1183

\bibitem[\protect\citeauthoryear{Kobulnicky et al.}{Kobulnicky, Kennicutt \& Pizagno}{1999}]{KKP99} 
Kobulnicky, H.A, Kennicutt, R.C.Jr. \& Pizagno, J.L. 1999, ApJ 514, 544
\bibitem[\protect\citeauthoryear{Kobulnicky \& Kewley}{Kobulnicky \& Kewley}{2004}]{KK04} 
Kobulnicky H.~A. \& Kewley L.~J. 2004, ApJ, 617, 240 
\bibitem[\protect\citeauthoryear{Kobulnicky et al.}{Kobulnicky et al.}{2003}]{Kobulnicky03}
Kobulnicky, H.~A., Willmer, C.N.A., Phillips, A.C., Koo, D.C., Faber, S.M., Weiner, B.J., Sarajedini, V.L., Simard, L. \& Vogt, N.P. 2003, ApJ 599, 1006 

\bibitem[\protect\citeauthoryear{Leitherer et al.}{Leitherer et al.}{1999}]{L99}
Leitherer, C., Schaerer, D., Goldader, J.D., Gonz\'alez-Delgado, R.M., Robert, C., Kune, D.F., de Mello, D.F., Devost, D. \& Heckman, T.M. 1999, ApJS, 123, 3 (\emph{STARBURST~99})

\bibitem[\protect\citeauthoryear{Lilly, Carollo \& Stockton}{Lilly et al.}{2003}]{Lilly03}
Lilly, S.J., Carollo, C.M. \& Stockton, A.N. 2003, ApJ, 597, 730

\bibitem[L\'opez-S\'anchez(2006)]{LS06}
L\'opez-S\'anchez, \'A.R. 2006, PhD Thesis, Universidad de la Laguna (Tenerife, Spain)
\bibitem[\protect\citeauthoryear{L\'opez-S\'anchez et al.}{L\'opez-S\'anchez, Esteban \& Rodr\'{\i}guez}{2004a}]{LSER04a}
L\'opez-S\'anchez, \'A.R., Esteban, C. \& Rodr\'{\i}guez, M. 2004a, ApJS, 153, 243
\bibitem[\protect\citeauthoryear{L\'opez-S\'anchez et al.}{L\'opez-S\'anchez, Esteban \& Rodr\'{\i}guez}{2004b}]{LSER04b}
L\'opez-S\'anchez, \'A.R., Esteban, C. \& Rodr\'{\i}guez, M. 2004b, A\&A, 428, 445
\bibitem[\protect\citeauthoryear{L\'opez-S\'anchez et al.}{L\'opez-S\'anchez, Esteban \& Garc\'{\i}a-Rojas}{2006}]{LSEGR06}
L\'opez-S\'anchez, \'A.R., Esteban, C. \& Garc\'{\i}a-Rojas, J. 2006, A\&A, 449, 997
\bibitem[\protect\citeauthoryear{L\'opez-S\'anchez et al.}{L\'opez-S\'anchez et al.}{2007}]{LSEGRPR07}
L\'opez-S\'anchez, \'A.R., Esteban, C., Garc\'{\i}a-Rojas, J., Peimbert, M. \& Rodr\'{\i}guez, M. 2007, ApJ, 656, 168
\bibitem[\protect\citeauthoryear{L\'opez-S\'anchez \& Esteban}{L\'opez-S\'anchez \& Esteban}{2008}]{LSE08} 
L\'opez-S\'anchez, \'A.R. \& Esteban, C. 2008, A\&A, 491, 131, Paper~I
\bibitem[\protect\citeauthoryear{L\'opez-S\'anchez \& Esteban}{L\'opez-S\'anchez \& Esteban}{2009}]{LSE09} 
L\'opez-S\'anchez, \'A.R. \& Esteban, C. 2009, A\&A, 508, 615, Paper~II
\bibitem[\protect\citeauthoryear{L\'opez-S\'anchez \& Esteban}{L\'opez-S\'anchez \& Esteban}{2010a}]{LSE10a} 
L\'opez-S\'anchez, \'A.R. \& Esteban, C. 2010a, A\&A, in press, Paper~III
\bibitem[\protect\citeauthoryear{L\'opez-S\'anchez \& Esteban}{L\'opez-S\'anchez \& Esteban}{2010b}]{LSE10b} 
L\'opez-S\'anchez, \'A.R. \& Esteban, C. 2010,b A\&A, in press, Paper~IV
\bibitem[\protect\citeauthoryear{L\'opez-S\'anchez et al.}{L\'opez-S\'anchez et al.}{2010}]{LS10} 
L\'opez-S\'anchez, \'A.R. et al. 2010, A\&A, in revision, Paper~V


\bibitem[\protect\citeauthoryear{McCall et al.}{McCall, Rybski \& Shields,}{1985}]{MRS85}
McCall, M.L., Rybski, P.M. \& Shields, G.A. 1985, ApJS 57, 1
\bibitem[\protect\citeauthoryear{McGaugh}{McGaugh}{1991}]{McGaugh91}
McGaugh, S.S. 1991, ApJ, 380, 140



\bibitem[\protect\citeauthoryear{Nagao et al.}{Nagao, Maiolino \& Marconi}{2006}]{Nagao06}
Nagao, T., Maiolino, R. \& Marconi, A. 2006, A\&A, 459, 85

\bibitem[\protect\citeauthoryear{Oey \& Shields}{Oey \& Shields}{2000}]{OS00}
Oey M.~S. \% Shields J.~C., 2000, ApJ, 539, 687 

\bibitem[\protect\citeauthoryear{Pagel et al.}{Pagel et al.}{1979}]{Pag79}
Pagel, B. E. J., Edmunds, M. G., Blackwell, D. E., Chun, M. S., Smith, G. 1979, MNRAS, 189, 95
\bibitem[\protect\citeauthoryear{Peimbert}{Peimbert}{1967}]{P67}
Peimbert, M. 1967, ApJ, 150, 825
\bibitem[\protect\citeauthoryear{Peimbert et al.}{Peimbert et al.}{2007}]{Peimbert07}
Peimbert, M., Peimbert, A.. Esteban, C.; Garc\'{\i}a-Rojas, J., Bresolin, F., Carigi, L., Ruiz, M.T. \& L\'opez-S\'anchez, \'A.R. 2007, RMxAC, 29, 72

\bibitem[\protect\citeauthoryear{P{\'e}rez-Montero \& D\'{\i}az}{P{\'e}rez-Montero \& D\'{\i}az}{2005}]{PerezMontero05}
P{\'e}rez-Montero, E. \& D{\'{\i}}az, A.~I.\ 2005, MNRAS, 361, 1063

\bibitem[\protect\citeauthoryear{Pettini \& Pagel}{Pettini \& Pagel}{2004}]{PP04}
Pettini, M. \& Pagel, B.E.J. 2004, \mnras, 348, 59
\bibitem[\protect\citeauthoryear{Pilyugin}{Pilyugin}{2000}]{P00}
Pilyugin, L.S. 2000, A\&A, 362, 325
\bibitem[\protect\citeauthoryear{Pilyugin}{Pilyugin}{2001a}]{P01a}
Pilyugin, L.S. 2001a, A\&A, 369, 594
\bibitem[\protect\citeauthoryear{Pilyugin}{Pilyugin}{2001b}]{P01b}
Pilyugin, L.S. 2001b, A\&A, 374, 412
\bibitem[\protect\citeauthoryear{Pilyugin et al.}{Pilyugin, V\'{\i}lchez \& Contini}{2004}]{P04}
Pilyugin, L.~S., V{\'{\i}}lchez, J.~M. \& Contini, T.\ 2004, A\&A, 425, 849 
\bibitem[\protect\citeauthoryear{Pilyugin \& Thuan}{Pilyugin \& Thuan}{2005}]{PT05}
Pilyugin, L.S. \& Thuan, T.X. 2005, ApJ, 631, 231

\bibitem[\protect\citeauthoryear{Schaerer et al.}{Schaerer, Contini \& Pindao}{1999}]{SCP99}
Schaerer, D., Contini, T. \& Pindao, M. 1999, A\&AS 136, 35

\bibitem[\protect\citeauthoryear{Stasi\'nska}{Stasi\'nska}{2002}]{Stasinska02}
Stasi\'nska, G. 2002, RMxAC, 12, 62
\bibitem[\protect\citeauthoryear{Stasi\'nska}{Stasi\'nska}{2005}]{Stasinska05}
Stasi\'nska, G. 2005, A\&A, 434, 507
\bibitem[\protect\citeauthoryear{Stasi\'nska}{Stasi\'nska}{2006}]{Sta06}
Stasi\'nska, G. 2006, A\&A, 454, 127
\bibitem[\protect\citeauthoryear{Stasi\'nska}{Stasi\'nska}{2009}]{Stasinska09}
Stasi\'nska, G. 2009, proceedings of IAU sumposium 262, \emph{Stellar Populations - planning for the next decade}, eds Bruzual \& Charlot, astro-ph:0910.0175
\bibitem[\protect\citeauthoryear{Storchi-Bergmann et al.}{Storchi-Bergmann, Calzetti \& Kinney}{1994}]{SBCK94}
Storchi-Bergmann, T., Calzetti, D. \& Kinney, A.L. 1994, ApJ, 429, 572
\bibitem[\protect\citeauthoryear{Sutherland \& Dopita}{Sutherland \& Dopita}{1993}]{SDopita93}
Sutherland, R.S. \& Dopita, M.A. 1993, ApJS, 88, 253


\bibitem[\protect\citeauthoryear{Teplitz et al.}{Teplitz et al.}{2000}]{Teplitz00}
Teplitz, H.I., Malkan, M.A., Steidel, C.C., McLean, I.S., Becklin, E.E., Figer, D.F., Gilbert, A.M., Graham, J.R., Larkin, J.E., Levenson, N.A. \& 
Wilcox, M.K. 2000, ApJ, 542, 18
\bibitem[\protect\citeauthoryear{Torres-Peimbert, Peimbert \& Fierro}{Torres-Peimbert et al.}{1989}]{Torres-Peimbert89}
Torres-Peimbert, S., Peimbert, M. \& Fierro, J. 1989, ApJ, 345, 186
\bibitem[\protect\citeauthoryear{Tremonti et al.}{Tremonti et al.}{2004}]{Tremonti04}
Tremonti, C.A., et al. 2004, ApJ, 613, 898

\bibitem[\protect\citeauthoryear{van Zee et al.}{van Zee, Salzer \& Haynes}{1998}]{vZee98}
van Zee, L., Salzer, J.J. \&  Haynes, M.P. 1998, ApJ, 497, 1
\bibitem[\protect\citeauthoryear{V\'azquez \& Leitherer}{V\'azquez \& Leitherer}{2005}]{VL05}
V\'azquez, G.A. \& Leitherer, C. 2005, ApJ, 621, 695
\bibitem[\protect\citeauthoryear{Vila-Costas \& Edmunds}{Vila-Costas \& Edmunds}{1992}]{Vila-Costas92}
Vila-Costas, M. B. \& Edmunds, M. G. 1993, MNRAS, 259, 121
\bibitem[\protect\citeauthoryear{Vila-Costas \& Edmunds}{Vila-Costas \& Edmunds}{1993}]{Vila-Costas93}
Vila-Costas, M. B. \& Edmunds, M. G. 1993, MNRAS, 265, 199
\bibitem[\protect\citeauthoryear{V\'{\i}lchez \& Esteban}{V\'{\i}lchez \& Esteban}{1996}]{Vil96}
V\'{\i}lchez, J.M., \& Esteban, C. 1996, MNRAS, 280, 720

\bibitem[\protect\citeauthoryear{Woosley \& Weaver}{Woosley \& Weaver}{1995}]{WoosleyWeaver95}
Woosley S.~E. \& Weaver, T.A. 1995, ApJS, 101, 181 
\bibitem[\protect\citeauthoryear{Yin et al}{Yin et al}{2007}]{Yin+07} 
Yin, S.Y., Liang, Y.C., Hammer, F., Brinchmann, J., Zhang, B., Deng, L.C. \& Flores, H., 2007, A\&A, 462, 535
\bibitem[\protect\citeauthoryear{York et al.}{York et al.}{2000}]{York00}
York, D.G. et al. 2000, AJ, 120, 1579

\bibitem[\protect\citeauthoryear{Zaritsky et al.}{Zaritsky, Kennicutt \& Huchra}{1994}]{ZKH94} 
Zaritsky, D., Kennicutt, R. C. Jr. \& Huchra, J.P. 1994, ApJ 420, 87



}

\bibliographystyle{aa} 
\end{thebibliography}
\end{document}